\documentclass[sigconf]{acmart}
\AtBeginDocument{%
  \providecommand\BibTeX{{%
    \normalfont B\kern-0.5em{\scshape i\kern-0.25em b}\kern-0.8em\TeX}}}

\setcopyright{acmcopyright}
\copyrightyear{2024}
\acmYear{2024}
\setcopyright{acmlicensed}\acmConference[CHI '24]{Proceedings of the CHI Conference on Human Factors in Computing Systems}{May 11--16, 2024}{Honolulu, HI, USA}
\acmBooktitle{Proceedings of the CHI Conference on Human Factors in Computing Systems (CHI '24), May 11--16, 2024, Honolulu, HI, USA}
\acmDOI{10.1145/3613904.3643043}
\acmISBN{979-8-4007-0330-0/24/05}

\usepackage{xcolor}

\usepackage{comment}

\newenvironment{quoteitalicized}
    {\begin{quote}}
    {\end{quote}}
\newcommand{\quotes}[2]{\begin{quoteitalicized}\textit{#1} -{#2}\end{quoteitalicized}}

\begin{document}

\title{Integrating measures of replicability into scholarly search: Challenges and opportunities}


\author{Chuhao Wu}
\affiliation{%
  \institution{The Pennsylvania State University}
  \city{State College}
  \state{Pennsylvania}
  \country{USA}
  \postcode{16802}
}
\email{cjw6297@psu.edu}

\author{Tatiana Chakravorti}
\affiliation{%
  \institution{The Pennsylvania State University}
  \city{State College}
  \state{Pennsylvania}
  \country{USA}
  \postcode{16802}
}
\email{tfc5416@psu.edu}

\author{John M. Carroll}
\affiliation{%
  \institution{The Pennsylvania State University}
  \city{State College}
  \state{Pennsylvania}
  \country{USA}
  \postcode{16802}
}
\email{jmc56@psu.edu}

\author{Sarah M. Rajtmajer}
\affiliation{%
  \institution{The Pennsylvania State University}
  \city{State College}
  \state{Pennsylvania}
  \country{USA}
  \postcode{16802}
}
\email{smr48@psu.edu}


\begin{abstract}
  Challenges to reproducibility and replicability have gained widespread attention, driven by large replication projects with lukewarm success rates. A nascent work has emerged developing algorithms to estimate the replicability of published findings. The current study explores ways in which AI-enabled signals of confidence in research might be integrated into the literature search. We interview 17 PhD researchers about their current processes for literature search and ask them to provide feedback on a replicability estimation tool. Our findings suggest that participants tend to confuse replicability with generalizability and related concepts. Information about replicability can support researchers throughout the research design processes. However, the use of AI estimation is debatable due to the lack of explainability and transparency. The ethical implications of AI-enabled confidence assessment must be further studied before such tools could be widely accepted. We discuss implications for the design of technological tools to support scholarly activities and advance replicability.
 
\end{abstract}




\maketitle

\section{Introduction}
\label{sec:Introduction}
Isaac Newton famously said, ``If I can see further, it is because I am standing on the shoulders of giants.'' The principle that scientific inquiry is built upon insights derived from an existing body of work is central to the scientific method \cite{cohen2011introduction}, and literature review has been proposed as the first and foundational step in any research project \cite{garfield_proposal_1977,baker_writing_2000}. Despite its important role, for many years, the lack of systematization challenges validity and undermines the reproducibility of findings based on meta-analyses 
\cite{harari_literature_2020,king_search_2020}. Several recommendations and guidelines have been proposed to help scholar report their literature search and review in a clear and reproducible manner. For example, the Preferred Reporting Items for Systematic Reviews and Meta-Analyses (PRISMA) \cite{page2021prisma} and its literature search extension \cite{rethlefsen2021prisma} have provided comprehensive checklists for reporting systematic reviews. However, besides methodological issues, there is also a technological component that needs to be addressed \cite{bethard_who_2010}, especially tools that can assess the quality of the results from search engines \cite{hinderks2020slr}. Systems for information retrieval, such as search engines, mainly rely on keyword matching, yet scientific article search ought to consider other factors such as citation patterns, publication venues, and disciplines. 

To tackle this problem, a number of information retrieval models and paper recommenders have been developed to incorporate citation context \cite{liu_literature_2014} and scholars' preferences \cite{zhu_recommending_2021} in their algorithm, and leverage text-mining to automate manual tasks and save time for researchers \cite{fortino2020using}. However, a comparative analysis of existing 
tools in the literature and quality metrics is needed to improve the usability, ease of use, reliability, performance, and support for future designs \cite{cestero2022pysurveillance}. \citet{al2019identification} show that researchers could have diverse needs for literature tools and there can be disciplinary differences. Yet, there is a lack of understanding of user experiences and challenges with these tools. For example, many researchers rely on digital databases and search engines such as Google Scholar for their literature search, but these digital libraries have their different features and may limit user experience and fail to meet user needs \cite{sosnicki2021ash}. As most of them are proprietary, we have little information about how users interact with and assess the search results. Understanding current practices around literature search can significantly contribute to the human-centered design of tools to support scholarly activities.

A primary challenge to the search and assembly of existing research is the massive and increasing number of publications across virtually all fields of study. The rise of predatory journals is flooding the literature with low-quality work \cite{bartholomew2014science,lalu2017stakeholders,sorokowski2017predatory}, heightening the stakes for effective search and, in parallel, motivating the inclusion of indicators of credibility into search outputs. Generative AI will likely exacerbate this problem in the coming years because of the potential misuse of AI in writing \cite{hill2023chat}.  
Even putting aside predatory journals and AI-generated papers, broader concerns about confidence in published work have come to the foreground over the past decade. 
Across the natural and social sciences, researchers have recognized that many studies are difficult or impossible to replicate, e.g., \cite{errington2014open,mcshane2019large}. A number of contributing factors have been proposed including publication bias favoring novel, affirmative results, manipulative statistical analyses, e.g.,  p-hacking, and lack of transparency when describing research methods \cite{munafo_manifesto_2017, baker_reproducibility_2016, collaboration_estimating_2015}. 

In response, the open science and science of science communities have made significant efforts to improve confidence in research, calling for changes to both policy and practice \cite{foster_open_2017,mckiernan_how_2016}. Likewise, technologies are emerging to automate aspects of confidence assessment for published findings. There have been a number of recent efforts, for example, using supervised learning over features extracted from papers' text and metadata to predict outcomes study replications \cite{altmejd2019predicting,yang2020estimating,pawel2020probabilistic,wu2021predicting, rajtmajer2021synthetic}. 

Yet, ways in which open science indicators, e.g., preregistration, open materials, as well as scores and explanations returned by automated approaches for confidence assessment, could be integrated into scholarly search and literature review, remain unexplored. To address this gap, we conduct semi-structured interviews with PhD students and postdoctoral researchers from social \& behavioral science departments in the U.S. We ask participants about their literature search and review practices, and about their concerns with respect to the credibility of work in their area. We also demonstrate an AI-driven replicability estimation tool to understand whether and how such a tool could assist them in their workflows. 
Specifically, our work is motivated by three primary research questions: 
\begin{itemize}
    \item RQ1: What are researchers' current approaches to literature search and review?
    \item RQ2: What challenges do researchers encounter during literature search and review?
    \item RQ3: How should signals of credibility be integrated into researchers' literature search and review?
\end{itemize}

It should be noted that there can be significant differences in the literature search process and perceptions of reproducibility across disciplines \cite{mengist2020method, freese2022advances, furlong2023toward}. Despite some common practices used regardless of discipline, the current study does not seek to make statements that can be generalized without further validation. As a context, this study only focuses on social and behavioral science research, with all participants and study materials coming from this field. With this qualification, the study makes several important contributions. First, we connect with and contribute to studies on design that support scholarly search and management by empirically documenting researchers' strategies for literature search, review, and evaluation, highlighting the need for more flexible and intelligent approaches. Second, we link researchers' literature review practices with their perceptions of credibility, demonstrating opportunities for the inclusion of reproducibility and replicability metrics. Finally, we explore scholars' perceptions of an AI-driven replicability estimation tool and how to integrate the replication prediction into the literature review.  We believe that the findings and discussions can motivate the research community to further ponder the design implications and ethical considerations for systems enabling automated assessment of confidence in published findings.

\section{Related work}
\label{sec:related}
Our work builds upon and brings together literature in the areas of scholarly search and scientific integrity and its assessment. In this section, we presented some prior progress made for augmenting literature search and reviewing, promoting research reproducibility and replicability, and assessing these dimensions of the quality of publications.

\subsection{Literature search and review}
\label{literaturesearch}
Effective literature search and review is foundational for research in all evidence-based fields, yet finding, assembling, and contextualizing relevant work is a painstaking and primarily \emph{ad hoc} manual process \cite{brocke_reconstructing_2009}. Researchers in various disciplines have attempted to organize protocols to guide effective and efficient literature search. Through the analysis of nine guidance documents on a systematic review of topics in the social sciences, \citet{cooper_defining_2018} summarize eight key stages of literature search, starting from ``who should literature search'' and ending with ``managing references and reporting the search process''. Other researchers have demonstrated that apparently slight differences in any stage of a literature search can lead to distinct results. For example, comparing search strategies used in 152 applied psychology systematic reviews, \citet{harari_literature_2020} highlight the important impact of database selection. They find that publisher databases tended to perform poorly with low yields, while Google Scholar returned the largest proportion of articles yet with poor precision. When using these databases, search queries also play an important role and sophisticated techniques may be required for users to construct an effective search query \cite{atkinson_how_2018,bramer_systematic_2018}.

Recognizing the challenges inherent to literature search and opportunities for technologies to provide assistance, Human-Computer Interaction (HCI) research has extensively explored ideas for supporting scholars in this work. \citet{choe_papers101_2021} design an interactive system (Papers101) to accelerate the discovery of literature by recommending relevant keywords and ranking papers based on keyword similarity, publication year, citation count, and other metrics. Action Science Explorer (ASE) \cite{dunne_rapid_2012} and LitSense \cite{sultanum_understanding_2020} support reference management and help researchers develop a holistic understanding of the literature through network visualizations, e.g., topic graphs and citation networks. While these papers have demonstrated the preliminary usability of their tools, evaluations were conducted with a limited, fixed set of publications. Therefore, they focused more on the effectiveness of visualizations and metrics rather than the whole search experience. Another line of work focuses on scaffolding scholarly reading and evaluation processes. Prior studies have developed novel interfaces to facilitate the understanding of technical terms and symbols used throughout a paper by providing tooltips and auto-generated glossaries \cite{head_augmenting_2021}, prioritizing and contextualizing inline citations based on researchers' reading history \cite{chang_citesee_2023}, and even augmenting documents with auto-generated paragraph headings and filtering redundant information \cite{palani_relatedly_2023}. With the advancement of natural language processing (NLP) techniques, we certainly expect continued innovation in support of the literature review. 

With the widespread popularity of scholarly search engines \cite{tenopir_seeking_2019}, understanding search practices is foundational to the design and development of these tools. \citet{maloney2016expecting} summarize four types of scholarly search strategies: \textit{Exploring}, \textit{Finding}, \textit{Refinding}, and \textit{Serendipity} based on whether users know what they are seeking and where to find it. They emphasize that information providers should particularly support serendipitous discovery. Through 368 survey responses, \citet{soufan2022searching} find that the literature search task is often identified as an exploratory search task characterized by unfamiliarity with the domain and dynamic information needs. As digital communication methods become diverse, literature search may also go beyond digital libraries and involve social resources and non-library technologies \cite{ince2018study}. It is clear that with the rapidly increasing volume of information, users will need more customized support from search tools since the optimal search strategy often depends on the task \cite{athukorala2016exploratory}. It is critical for search tools to organize results for the optimal usefulness of the information to the users \cite{wang2007learn}. Therefore, understanding how users assess search outcomes plays an important role in improving the design, as does understanding the ways in which current systems satisfy or fail to meet researchers' needs.

\subsection{The replication crisis}
\label{replication}
Another important object of the study is to explore how signals of replicability could be integrated into researchers' literature search and review process. Concerns have been raised over the past decade about the reproducibility, replicability, and robustness of published findings in the social sciences and beyond have gained significant attention, as large-scale efforts to replicate high-profile empirical studies have yielded low success rate \cite{camerer_evaluating_2016,collaboration_estimating_2015}. This revelation has been nicknamed the \emph{replication crisis}, or equivalently the \emph{reproducibility crisis}. There has been some ambiguity around terms; throughout this work, we adopt definitions from \cite{national2019reproducibility,nosek2021replicability,pineau2021improving}. Namely, \emph{reproducibility} refers to computational repeatability – obtaining consistent computational results using the same data, methods, code, and conditions of analysis; \emph{replicability} means obtaining consistent results on a new dataset using similar methods. These two terms have been frequently mentioned together or even used interchangeably due to their close relatedness \cite{milkowski_replicability_2018,plesser_reproducibility_2018}. \emph{Robustness} is obtaining consistent results on the same data using a different analytical approach. \emph{Generalizability}
refers to the extent that results hold in other contexts or populations different from the original. Each of these terms captures some subset of qualities we might consider important for contextualizing confidence in a given claim or finding. 

\citet{baker_reproducibility_2016} showed there was a crisis of reproducibility due to selective reporting, low statistical power, and other inappropriate practices. Other studies have confirmed that questionable research practices (QRPs) such as p-hacking and HARKing (constructing new hypotheses after the results are known) can yield false-positive results \cite{head_extent_2015}, and journals are less likely to publish replication studies, especially those contradicting prior findings \cite{makel_replications_2012}. Recognizing this replication crisis, the scientific community has proposed multiple initiatives to address these issues. For example, making study materials such as data and analysis codes openly available allows others to validate published results and identify errors more easily \cite{kidwell_badges_2016}. Platforms such as Open Science Framework (OSF) \cite{foster_open_2017} also facilitated data sharing and promoted reproducible practices across the community. 

While researchers' conscientiousness in maintaining good practices is critical, incentive and policy changes are also necessary. Despite exciting progress in both promoting and assessing reproducibility and replicability, to the best of our knowledge, there has been limited study of how these considerations should be integrated into scholarly search and review. Specifically, it is unclear whether researchers look for signals of reproducibility during their search processes, and to what extent this criterion may impact their selection of literature. Considering the significant role of literature review in the research workflow, the current study explores this issue by focusing on whether and how researchers would include automated assessments of the replicability of the findings in their workflows.

\subsection{Assessment of reproducibility and replicability}
\label{assessment}

A number of large-scale replication projects have set out to assess the state of reproducibility and replicability in specific domains.  These have included efforts targeting psychology \cite{open2015estimating}, economics \cite{camerer2016evaluating,berry2017assessing}, experimental philosophy \cite{cova2021estimating}, social and behavioral sciences more broadly \cite{camerer2018evaluating}, cancer biology \cite{errington2014open}, neuroscience \cite{botvinik2020variability}, computer science \cite{collberg2014measuring}, and machine learning \cite{raff2019step}. Generally speaking, these projects require massive investment in time and resources and, therefore necessarily limited in scale. The most common approach to assessment of reproducibility and replicability, short of actually repeating a study, is to apply a checklist of evaluation criteria to a paper in question. For example, to estimate the state of reproducibility of AI research, \citet{gundersen_state_2018} use a list of binary variables (true or false) indicating how well the method, data, and experiment are documented in a paper. Similarly, for web measurement studies, \citet{demir_reproducibility_2022} specify 18 criteria with three levels (omit, undocumented, and satisfy), describing the dataset, experiment design, and evaluation aspects of a paper. However, these checklists still require manual examination of the full paper. In recent years, researchers have begun to explore opportunities for automated estimation of replicability, e.g., \cite{altmejd2019predicting,pawel2020probabilistic,yang2020estimating} using supervised learning over features extracted from text and metadata of ground truth replication project outcomes. While these applications are promising, there are several important issues known in machine learning and AI-empowered tools that might impact user experience. As \citet{chen2022evaluating} suggested, there is an increasing need for search engines, recommenders, and similar systems to provide transparent and explainable results to users. Existing research on explainable AI (XAI) emphasized that systems should enable users to see how inputs and outputs are mathematically interlinked, using an intrinsic method that generates a human-readable explanation for the model’s decision \cite{ali2023explainable}. Some researchers pointed out that previous approaches to XAI have been algorithm-centered, and understanding end-users' explainability needs and the socio-organizational context should be incorporated in the development of XAI \cite{ehsan2021expanding, kim2023help}. In addition, for decision-making AI systems, robust representation of uncertainty also plays an essential role during human-AI collaboration \cite{cassenti2021robust,prabhudesai2023understanding}. Therefore, examining user perception of explainability, transparency, and uncertainty is critical for designing automated estimation of replicability.

\section{AI prediction market for research confidence assessment}
\label{AImarket}
The current study takes advantage of an AI-driven tool for confidence scoring the replicability of published findings \cite{rajtmajer2021synthetic}. The inner workings of the AI are powered by a synthetic prediction market, i.e., a market where trader bots are asked to buy and sell outcomes of a notional replication study for a given finding. The feature extraction framework for replicability prediction is designed to extract a comprehensive range of features from published papers and their metadata, spanning five distinct categories: bibliometric, venue-related, author-related, statistical, and semantic information. Currently, it extracts a total of 41 features, encompassing various elements such as p-values, sample size, the number of authors, and the acknowledgment of funding. This broad array of features provides a detailed perspective on the papers, aiding in the assessment of their replicability. In the market, agents are initialized with a set of extracted features that represent the paper. 
The base model of the artificial binary prediction market is described in \cite{nakshatri2021design}. In the artificial prediction markets, AI agents are trained using a genetic algorithm, where agents that generate profits are retained while those failing to do so are removed from the agent pool. During this process, the five most profitable agents are identified, from which the top three are selected for mutation and crossover. These agents are endowed with a predetermined amount of initial cash, enabling them to purchase assets. The logic governing these purchases is defined by a sigmoid transformation function, with the stipulation that agents may acquire only one share of an asset at a time. The genetic algorithm's objective function aims to minimize the root mean square error (RMSE) of the estimated score. The aggregate final market price serves as the score value for each paper, essentially representing the market's final price. Various hyperparameters, such as the liquidity constant, market duration, initial cash, number of generations, number of agents per market, agent inter-arrival rate, and run time, play a pivotal role in controlling market performance. The careful selection of these hyperparameters is crucial for optimal market functionality \cite{wu2021predicting,rajtmajer2022synthetic}. Since agents are fully algorithmic, similar to neurons in a neural network, we do not need to have ground truth for test data points (research findings). Agents are, however, trained on 446 ground truth replication outcomes.

\section{Ethical considerations}
The prediction tool described above is fully developed by prior research \cite{rajtmajer2021synthetic}. It was obtained with full consent from the corresponding author. The current study does not make any modifications or design decisions for the system. During our interviews, we only demonstrate this tool to participants and collect feedback. Our overarching aim is to examine whether and how technological assistance (whether via the tool or some other means) can facilitate researchers' estimation of confidence in published work for the purposes of literature review and synthesis. Nonetheless, we need to point out some important ethical considerations for such a tool. First of all, regarding the use of author-related information, the tool only utilizes information presented in a publication (e.g., affiliations). It does not collect or use any extra demographic information about the authors (e.g., age and gender). Second, we acknowledge that researchers may not be comfortable with AI-empowered replicability estimation, as AI predictions are often based on correlations instead of causality \cite{ganguly2023review}. Therefore, to what extent these tools accurately estimate replicability is still uncertain and requires thorough examination. In addition, the sole focus on replicability does not necessarily improve the quality of research \cite{devezer2019scientific}. Therefore, the use of AI-empowered replicability estimation should be treated with extra caution. For this exploratory study, we aim to take a neutral position without enforcing certain attitudes on our participants. Instead, we investigate participants' perceptions of the issue and openly invite them to comment on the tool's negative and positive aspects.

\section{Methods}
\label{method}
\subsection{Recruitment and Data Collection}
\label{recruit}

\begin{figure*}[h] 
    \centering 
    \includegraphics[width=0.6\linewidth]{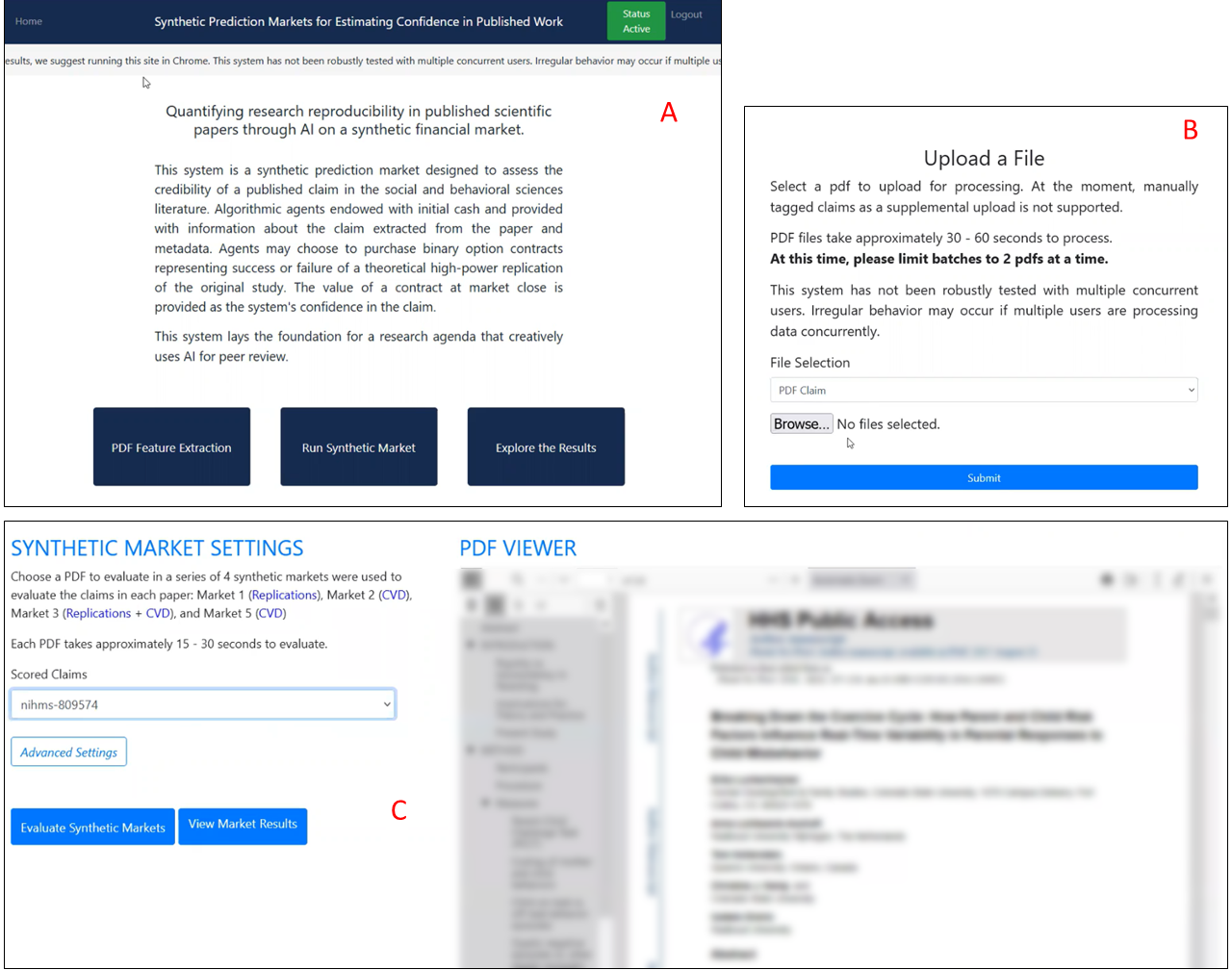} 
    \caption{Web interface for replicability estimation prototype tool. A: Home page; B: Interface for uploading PDFs for scoring; C: Interface for paper evaluation, with the ability for users to adjust model hyperparameters using Advanced Settings.}
    \label{fig:interface} 
\end{figure*}

\begin{figure*}[h] 
    \centering 
    \includegraphics[width=0.8\linewidth]{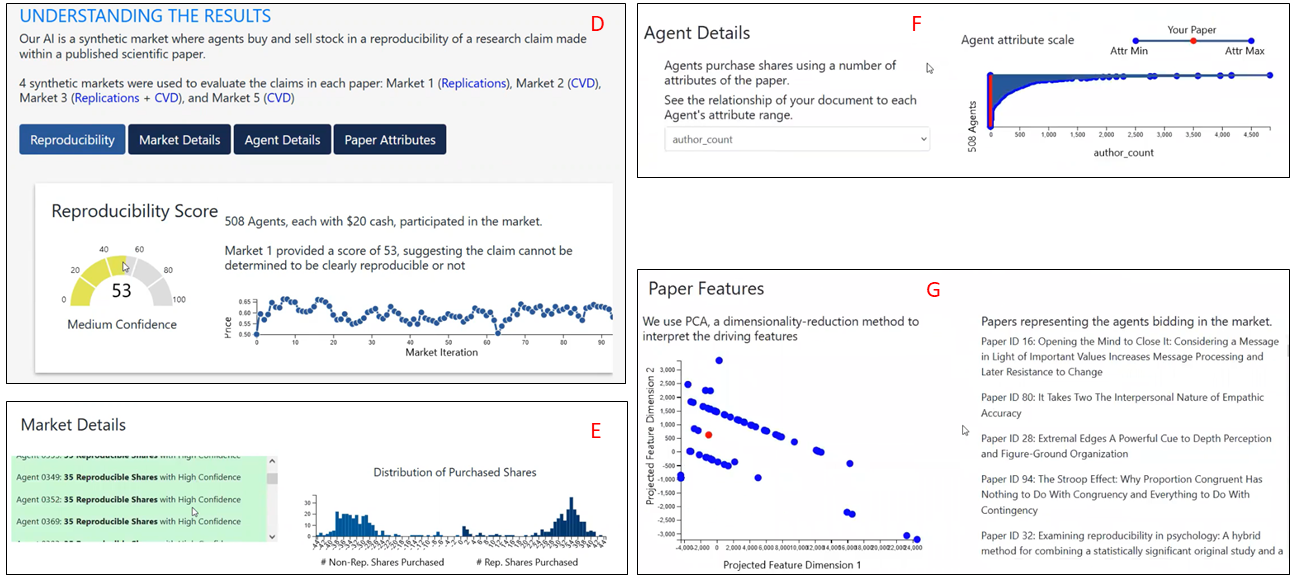} 
    \caption{Example display of paper evaluation results. D: Replicability score. E: Explanations, presented as details about agents' behaviors/decisions; F: Publication's features in context; G: Visualization of principal components from extracted features.}
    \label{fig:results}
\end{figure*}

The study took place on the main campus of an R1 university in the mid-Atlantic region of the United States. Participants were selected using purposive sampling. Recruitment emails were sent to PhD students currently enrolled in, or who had recently graduated from, social and behavioral science (SBS) disciplines at the university—such as psychology, sociology, and anthropology programs. This specific focus was chosen because discussions around the replicability crisis have been particularly prominent in SBS fields, and the AI prototype was trained on publications from these disciplines. However, it should be noted that the resulting findings will, therefore, be confined by the characteristics of the chosen population and should not generalized to the other disciplines and cultures without further validation.

Participants' email addresses were sourced from the university's directory, and the study received approval from the institution's Institutional Review Board (IRB). Data collection occurred from November 2022 to May 2023. The recruitment email included a hyperlink, enabling participants to voluntarily schedule remote interviews. Prior to the interview, participants were asked to complete a demographic questionnaire and provide the title of a recently-read paper. Interviews were conducted and recorded via Zoom video conferencing. For the first part of the interview, questions are centered around participants' experiences with literature search and assessment. In the second part, we demonstrated the prototype interface \cite{rajtmajer2021synthetic} using the paper chosen by the participant. The web-based interface comprises three main functions: PDF Feature Extraction, Run Synthetic Market, and Explore Results (See Figure~\ref{fig:interface}). In PDF Feature Extraction, users can upload PDFs of published papers for assessment, and the system automatically extracts relevant features. The Run Synthetic Market function then generates replicability estimates. Figure~\ref{fig:results} displays an example of these estimates, which range from 0 to 100; higher scores indicate a greater likelihood of successful replication. This score corresponds to the final price in the synthetic market. Users can delve deeper into the results by examining the decisions of individual agents (Market Details), comparing the paper's features against agent value ranges (Agent Details), and scrutinizing the principal components extracted from the features (Paper Features).

\begin{figure*}[h] 
    \centering 
    \includegraphics[width=0.8\linewidth]{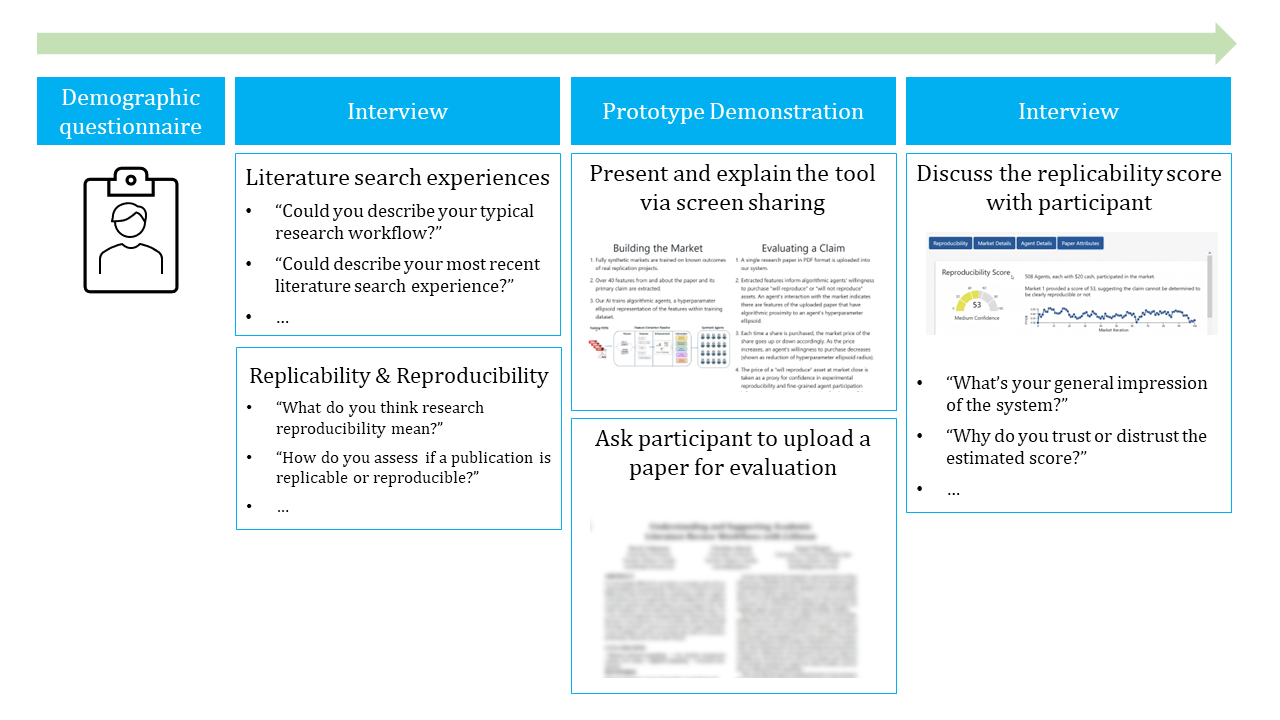}
    \caption{Illustration of the study protocol.}
    \label{fig:protocol}
\end{figure*}

This demonstration was followed by a series of questions exploring whether and how the replicability score might be useful in participants' research workflows. Additional questions were incorporated based on participants' responses during the interview. Figure~\ref{fig:protocol} outlines the study protocol. All data were collected with informed consent from the participants and subsequently anonymized. Interview recordings were transcribed for further analysis.

\textbf{Understanding of reproducibility.} During the interview, we asked participants about their understanding of reproducibility and provided relevant explanations if participants were not able to answer the question. However, we did not assess whether participants' understanding aligns with the strict definition. This study seeks to report participants' understanding of reproducibility and how they assess it in their daily practices. Therefore, correcting their understanding may lead to biased reflections on their previous experiences. In addition, the replicability score is estimated using various publications features instead of the definition of the concept. Following participants' current understanding of reproducibility enables us to obtain reactions to the system that align more closely with participants' needs and habits.

\subsection{Data analysis}
\label{dataanalyis}
Interview transcripts were analyzed using a thematic analysis approach \cite{braun2006using}. We adopted a collaborative and iterative coding process widely used by qualitative studies in HCI \cite{ammari2015understanding,huang2020you,ding2022uploaders} to ensure the validity of the analysis process. Specifically, the first author thoroughly read transcripts multiple times to become familiar with the data. In the subsequent phase, open coding was performed to identify initial codes that represent meaningful segments that captured participants' opinions and behavioral patterns. These initial codes were then organized into potential themes pertinent to the primary research questions. After this initial coding phase, the first author met with all co-authors to discuss the meaning, similarities, and differences of the identified themes. Decisions were made collectively by the team regarding the retention, removal, or reorganization of these themes. For example, we grouped codes related to the author's reputation, journal reputation, and citation number under the theme ``Metrics for filtering search results'', as these qualities are frequently used by participants to select papers in the literature search. Subsequently, the first author continually reviewed and refined the themes by revisiting the transcripts and confirming the connections between themes and the high-level research questions. Once the themes were finalized, the team met again to interpret and develop them into narratives shown in Section \ref{findings}. Analytic memos were employed throughout the process to supplement coding and facilitate reflections \cite{creswell2000determining}. Analysis was done through NVivo 12, a qualitative data analysis software. The data analysis process was done iteratively by moving back and forth among the transcripts. All authors are active researchers in social science and closely related fields, including HCI. Our extensive experiences in literature review and management enable the team to make sense of the interview data. While we aim to avoid biases and stay neutral during data collection and interpretation, we acknowledge that findings could be subjective and depend on the authors' understanding.

\section{Findings}
\label{findings}
Seventeen participants were interviewed, with an average interview length of 42 minutes. Participants' self-reported demographic information is shown in Table \ref{tab:subject}, including gender, ethnicity, PhD progress (the number of years), and their main research field. 

\begin{table}[h]
\caption{Participants' demographic information.}
\label{tab:subject}
\resizebox{\columnwidth}{!}{%
\begin{tabular}{|l|l|l|l|l|}
\hline
\multicolumn{1}{|c|}{\textbf{ID}} & \multicolumn{1}{c|}{\textbf{Gender}} & \multicolumn{1}{c|}{\textbf{Ethnicity}} & \multicolumn{1}{c|}{\textbf{PhD progress}} & \multicolumn{1}{c|}{\textbf{Research field}} \\ \hline
P1                                & Female                               & Asian                                   & 1                                          & Human-Computer Interaction                   \\
P2                                & Male                                 & White                                   & 1                                          & Management                                   \\
P3                                & Female                               & Asian                                   & 1.5                                        & Communications                               \\
P4                                & Female                               & White                                   & 5                                          & Communications                               \\
P5                                & Male                                 & Asian                                   & 2                                          & Management                                   \\
P6                                & Female                               & Pacific Islander                        & 3                                          & Marketing                                    \\
P7                                & Female                               & White                                   & 1                                          & Psychology                                   \\
P8                                & Female                               & White                                   & Postdoc                                    & Psychology                                   \\
P9                                & Male                                 & White                                   & 3.5                                        & Psychology                                   \\
P10                               & Female                               & Other                                   & 3                                          & Psychology                                   \\
P11                               & Female                               & White                                   & Postdoc                                    & Psychology                                   \\
P12                               & Male                                 & White                                   & 3.5                                        & Criminology                                  \\
P13                               & Female                               & White                                   & 3                                          & Sociology                                    \\
P14                               & Male                                 & Other                                   & 4                                          & Political Science                            \\
P15                               & Female                               & Asian                                   & 3                                          & Anthropology                                 \\
P16                               & Male                                 & Black                                   & 1                                          & Anthropology                                 \\
P17                               & Male                                 & Asian                                   & 3                                          & Human-Computer Interaction                   \\ \hline
\end{tabular}%
}
\end{table}

\subsection{Literature search: Strategies and challenges}
\label{strategies}
\subsubsection{Metrics for filtering search results}
\label{metrics}
Literature search constitutes a critical step in a researcher's workflow, particularly during the early stages of a research project.   Participants indicated that reviewing literature helped them to understand {``what is still unknown''}, check {``whether my questions are already answered''}, and establish {``the empirical basis for the idea and the theoretical rationale for why I want to test the question''}. 

Google Scholar emerges as the most commonly used search tool. Participants also utilized other platforms such as Web of Science and discipline-specific databases like PsycINFO. Although platforms like Web of Science offer advanced search queries and export functionalities, participants generally favored Google Scholar for its user-friendly interface. However, they noted that Google Scholar often yields overly broad and discipline-agnostic results. P6 mentioned that this issue diminished as her field expertise grew:

\quotes{I had this question when I first entered the PhD program: basically, where do you find the most important literature? But I guess after years of experiments, I now purely rely on Google Scholar, because I can already know, what are the important journals? And what are the reliable sources? I can distinguish the most important and relevant information from Google Scholar [results].}{P6}

As P6 noted, the choice of publication venue serves as a critical metric for evaluating search results. Other criteria include the number of citations, the author's reputation, and the recency of the publication. For instance, P5 values the reputation of authors or their institutions as indicators of research quality:

\quotes{It depends on the quality of the journal and impact factor, and depends on if I know the authors because I'm mostly familiar with all the prominent authors in my area. If I recognize the authors or the institute that the paper is from. Then I do deeper [reading]. If it is from Harvard, Stanford, Wharton, Michigan, I know that it's of top quality.}{P5}

Researchers new to a field, who are less familiar with established names, may rely more on citation counts as a measure of credibility:

\quotes{ I've studied psychology for about 10 years but as for Anthropology, I'm a newbie. So even if I see an author's name, there's a high chance I will not recognize it, even though he or she is a figure in the field. So, I have to rely on peripheral information like citation numbers. I would like to have at least two digits, even if it was published recently.}{P15}

While P15 held a high standard of requiring at least two digits in the citation count, P8 was less persuaded by the impact of the citation number:

\quotes{It's good if it's published in peer-reviewed journals and I don't tend to care about the impact factor or like how many citations.}{P8}

P8 contended that the peer review process alone validates research quality. The majority of participants, however, paid heed to additional metrics. Author reputation not only influenced the evaluation of search results but also aided in the discovery of further relevant papers. Participants would follow authors' profiles on Google Scholar and institutional websites to find publications of potential interest:

\quotes{Once I start to identify particular researchers who seem to be hitting on the topics that I am more interested in, I will go to Google the profiles of those researchers or look up their professional profiles from their institutions and try to read other stuff that they've done}{P7}

Participants also employed citation networks to identify relevant literature. For instance, after identifying a foundational paper, they may explore its cited references or papers that have cited it to find related works. Instead of research quality, a citation network is mainly helpful in ensuring that the search is related to a specific topic:

\quotes{One thing I'll do is I'll look up a foundational kind of article; pop it into Google Scholar and see who else cited it, and then see if there are any papers that are similar to the topic I want to go down.}{P4}

Publication recency is another commonly used metric, particularly for empirical or experimental studies. The age of a paper is generally less critical for theoretical contributions:

\quotes{Usually, I prefer a journal publication that was less than 10 years, but if I need to find some theory papers, I don't mind what year this paper was published.}{P1}

P7 also added that it would be fine for review and meta-analysis papers to be a little bit old, because \textit{``it'll still give me a summary of a lot of things and conceptualize thought pieces.''} P4 stated that it's important to have a mixture of recent publications and foundational papers that may be very old, which can \textit{``make sure that you are touching on those foundational pieces''}. Recency could also interact with other metrics during the assessment. For example, P11 would make exceptions for ``outdated'' papers if they were published by prominent researchers he could recognize. P10 pointed out the citation number may not be a valid indicator of quality as new papers tend to get fewer citations but can still be of good quality.

Despite some nuances in their importance, these easily accessible paper metrics were frequently used, and they helped participants quickly filter through the long list of search results. Still, these metrics provide limited information about a paper, and participants need to take further steps to assess papers and seek information useful to their own research.

\subsubsection{Paper evaluation and information seeking}
\label{sec:evaluation}
Evaluating papers in detail often commences with the abstract, which serves as \textit{``the synopsis of the article''} for most participants. Once the abstract confirms the paper's relevance, the participants' approaches to reading the remaining content diverge. P7, for instance, prioritized the results section because the data allowed her to see if hypotheses were supported and then interpret the statistics in a relatively independent manner:

\quotes{I want to see the general summary statistics that they can provide. Descriptive statistics and correlations and things. Just to give me the first bird's eye view of what might be going on with their data. And then I will read the hypotheses that they proposed beforehand and experiment... I prefer to look at the tables and the charts first, so that I can kind of draw my own conclusions and then see how they're presenting their information.}{P7}

P7 believed that the numbers were important for understanding how a paper's conclusions were formulated. It seemed that participants who prioritized the result section were more sensitive to numbers and statistics, as P10 commented:

\quotes{I have a lot of statistical background. So I kind of examine whether the statistics they chose match the question they were asking. And if I can follow why they chose the model, and if the results of that model seemed to speak to a meaningful finding, as opposed to maybe just like p-hacking. Those are the things I focus on.}{P10}

By contrast, P3 believed that it wouldn't be helpful to focus on the results due to the publication bias favoring positive or statistically significant results:

\quotes{It's almost never the results. Result is usually not the way that I filter. And part of that is the skepticism that for most papers, if it's hypothesized, then it will get published if the hypotheses are verified.}{P3}

Similarly, P16 pointed out it was the method section that determined the integrity of the paper, instead of the results:
\quotes{From the methods, you can know how detailed the sampling was. What was ethical compliance? And how they adjusted their methods for the research questions? So, for me, it's not really about the results, but it's about the method.}{P16}

P13 also emphasized other sections over results, but due to the potential complexity in understanding the details. Instead, he focused on discussion and conclusion to understand a paper:

\quotes{After the abstract, I'd say the intro, and then the discussion and conclusion. And then I usually go back and just quickly skim the literature review. Depending on what I'm trying to do, sometimes I'll read the data and methods as well. I feel like the results sometimes are just, there's too much going on here. So I'll just stick to the conclusion.}{P13}

Despite the differences, participant' priorities in paper evaluation often depend on their current research phase or specific informational needs. For example, despite her emphasis on results, P7 would scrutinize the methods section when venturing into an unfamiliar discipline:

\quotes{It really depends on what I was looking for. For example, if I don't know how to design the experiments for this specific type of study, I go to method part and also result part... If it's an experiment that's outside of my discipline, I look at the methods really heavily, because there are different standards and expectations in different disciplines.}{P7}

When dealing with unfamiliar topics, P10 would instead read the introduction carefully. Similarly, P4 would pay more attention to the early sections of a paper when learning a new research topic:


\quotes{It depends on where I'm at in my literature review. If it is a new topic, I'll be trying to figure out how people are defining certain concepts, then I probably do pay more attention to the literature review, just so I can figure out who are the big names in this field? What are the definitions?}{P4}

The introduction and literature sections were helpful for participants to figure out the main topics and theories discussed and their definitions, whereas the method section can inspire new ideas for experiment designs. As P6 summarized, the iterative nature of research often prompts revisiting papers with a changing focus, depending on the project's stage:

\quotes{I think that will be dependent on my goals. For example, when you are doing a literature review on the earliest stage, you may focus on what topics they cover, but as you proceed into the specific research question, you would probably look but more into the details of the specific methods they use. And at the end, when you try to write a paper then probably you will follow others' story lines and how they how they develop the whole paper. So I guess I will revisit a paper again from time to time during the whole project.}{P6}

Overall, participants employ utility-oriented strategies for evaluating papers, focusing on aspects they believe would either effectively assess a paper's quality or contribute to their own research. As P9 commented, he needed to do a lot of research, and time efficiency really mattered in the literature search.

\subsubsection{Challenges to literature search}
\label{challenes}
Participants primarily used keyword-based searches to locate literature, but this approach often presented challenges. They found that search engines like Google Scholar did not always yield results consistent with their keyword expectations. This inconsistency was especially prevalent in interdisciplinary or innovative research, where the same concept may have multiple names:

\quotes{My area of research is so interdisciplinary. Researchers in this field that I'm not familiar with, they might be studying this thing, and they are just calling it something that I wouldn't expect them to call it. So I feel like there is a level of randomness added to how I can discover papers on topics that I want to study. I just need to know what words to use. That's probably been my biggest challenge.}{P7}

To address this, P4 suggested that search engines should offer keyword recommendations akin to a thesaurus. P6, however, was skeptical of recommendation systems, citing her experience with irrelevant alerts from Google Scholar. She proposed that a visual representation of the citation network would be more beneficial. In fact, several participants echoed the need for enhanced visualization features, such as a graphical representation of citations or co-authorship networks. While Google Scholar fulfills part of the expectations by providing the list of ``Cited by'' papers and the list of ``Co-authors'', participants desired more advanced functionalities:

\quotes{Like a platform that you can give them a topic, and then it gives you the main authors that have published stuff related to that topic... I guess it could use metrics like the h-index or the number of citations, and I guess you can also use more subjective things like where do they work, or what journals they have published, or their academic ranking?}{P14}

Despite options like date ranges and author filters, current search methods often fall short of capturing the nuanced information that researchers seek, e.g., theoretical framework, research methods, and main findings. P13 likened this to online shopping experiences, expressing a desire for more customized search criteria:

\quotes{Almost like when you're shopping online: I want this size in this color. So this would be like, I want [papers that used] these types of dataset, [from] this time period, [from] these countries or regions even... The general method that they use, like if they're using like, quantitative methods versus qualitative methods.}{P13}

This desire for more specific details also reflected participants' need for more intelligent search tools that can summarize scientific reports. While abstracts are designed for quick navigation of a paper, participants have found them ``not standardized'' or ``too big picture''. Instead, an intelligent tool that could offer more tailored summaries could address this gap:

\quotes{If there was a way to click on a paper and see what survey items are used to measure each construct, to pull that out of a paper more automatically would be helpful. Along with it, results and directionality of results. If it's positively correlated, or negatively correlated with other things.}{P3}

In summary, while existing search methodologies offer a baseline utility, there is a growing need for more sophisticated, flexible, and intelligent tools. These should accommodate not only standardized keywords but also offer advanced features like network visualization and content-specific search options.

\subsection{Reproducibility and replicability: Importance and estimation}
\label{reproducibility}
\subsubsection{Perception of the importance}
\label{perception}

Except for P2 and P17, all participants reported that they've heard of the concept of research replicability or reproducibility. During the interview, many participants have used the terms interchangeably. Accordingly, we do not strictly distinguish between the two in the following sections.

Participants generally agreed that replication and reproduction of studies strengthen the credibility of scientific findings:

\quotes{A lot of how we evaluate statistical effects in psychology is p-value, which is the assessment of how likely these results are due to chance. So any single finding can always be due to chance. And so if you're reproducing studies and if you find the same results, it gives you a lot more confidence that it's probably not due to chance.}{P7}

Despite the emphasis on replication, P7's interpretation of p-values was incomplete and, therefore, biased his understanding of replicability. Specifically, a p-value actually measures the probability of observing the data, or something more extreme, if the null hypothesis is true \cite{colquhoun2017reproducibility}. The emphasis on `chance' also oversimplifies the size of the effect, the precision of the estimates, and the quality of the studies. Similarly, other participants highlighted some relevant aspects that are not at the core of the replicability crisis, demonstrating a lack of a strict definition of replicability but rather a general understanding of qualities for good research. For instance, while P7 and P14 noted that emphasizing replicability minimizes errors due to chance and guards against research misconduct, P8 was more concerned with the representativeness of study populations, emphasizing the need for diversity to enhance generalizability:
\quotes{I think a lot of our research has been based on what we call the WEIRD sample, which is Western, European, Industrialized [Rich, and Democratic population], you know. And I'm not surprised as our samples become more diverse in various ways, we can't find the same results. I think it's a distinct sample where a lot of psychology is based off, and a lot of our theories are based off... That could be one issue for the reproducibility crisis.}{P8}

Generalizability is often highly related to replicability in that non-replicable studies are typically considered non-generalizable. Yet, a study may be replicable but not generalizable if the findings only apply under specific conditions \cite{national2019reproducibility}. It seemed that participants conflated concerns about replicability with generalizability to some extent. For example, for certain types of research, participants regarded replicability as less important or not applicable, especially when the research is qualitative in nature or involves the context of human interaction:

\quotes{If you're doing interviews, it's impossible [to reproduce]. If someone were to interview me in a similar way that we're talking right now, but two weeks later, those interviews will be different. Even if someone was to just listen in our recorded interview. They still would miss the interaction that we're having.}{P3}

\quotes{I'm studying Organizational Psychology. And you cannot create the natural environments to understand people's behavior. In my field, context is the determinant. That's why I would understand that people cannot replicate studies}{P9}

In addition, participants noted that even though replicability is important in many fields, the effort of validating studies is much less appreciated by scholars than innovative findings:
\quotes{I've gotten reviews that say, it's important that this research (replication studies) is being done but the journals don't want to or not always interested in and publishing such work, because it doesn't advance the field forward, it only makes sure that the base that we have is solid. So while they appreciate the effort, they don't always want to publish it within the top tier journals.}{P12}

Their understanding of replicability, while not accurate, may reflect some common perceptions among researchers, highlighting the need for curriculum interventions. However, this mixture of replicability and other properties of research quality does not equal misunderstanding and has, nevertheless, motivated participants to pay attention to good research practices. For example, during the literature search, while platforms lack explicit metrics for reproducibility and replicability, participants have made assessments of these qualities to various extents when reading papers. As mentioned in Section \ref{sec:evaluation}, many participants paid significant attention to the method section of a paper, and by examining the rationale of the research methods, they were implicitly evaluating the replicability:
\quotes{I am a bit skeptical while reading the design and methods because I saw they hypothesize something and they are measuring in a way that has many flaws. I would say it's not sure if they can replicate this study under the same conditions.}{P9}

Again, this evaluation often included other aspects of the paper quality, such as generalizability:
\quotes{I would be paying attention to things like how generalizable is the sample that they recruit? How many people did they recruit? A lot of studies, especially older studies, had like teeny tiny little samples, and that may not be valuable. And then assuming it's a method I'm really familiar with, like eye tracking, or imaging, I can use that section to evaluate what they did was accurate as opposed to poorly followed methodology.}{P10}

More explicitly, participants said that clear descriptions of research methods, good rationale for research choices, and availability of research materials would convince them that a study is reproducible:

\quotes{I am all for a very clear procedure, stimuli and measure section. For me, that is important. I feel like a lot of times, authors say if you want to see the stimuli, email us, if you want the full list of measures, email us, I think that's fine. But especially as a more junior researcher, it is helpful to be able to see the step-by-step [methods].}{P4}

\quotes{I think it is (reproducible), because it's a widely available dataset, and has very clear instructions. There's not a lot of wiggle room in that.}{P8}

Some participants also strongly advocated for the practice of pre-registration. Pre-registration can ensure that a publication contains sufficient detail to permit reproduction and can reduce the bias in reporting significant results:  

\quotes{Ultimately, I really like the idea of the registered reports, because the journal is saying they're going to accept the article whether the results are null or not. So I think that's something I would want to focus on moving forward.}{P10}

Generally, participants acknowledged that reproducibility and replicability are important concepts for the research community. Despite some confusion between replicability and other aspects of research quality, participants have paid attention to reproducible practices during their literature search. Rigorous research methods and availability of research materials were the main proxies that enabled participants to estimate paper quality and thereby impact their literature review process.  

\subsubsection{Trust in AI-estimated replicability}
\label{trust}
The AI-estimated score offers a quantitative metric for assessing the replicability of research without the need for actual replication. Some participants felt the score aligned well with their impression of the specific paper in question. However, others were hesitant to trust the AI system without understanding its underlying mechanisms:

\quotes{For now, I don't think I trust it because I don't really understand how this system works. But once I understand how it work, and I think it's reasonable, I think I can trust more. A little bit of explanation of how these algorithms have been trained, and how many research papers are included in the training sets?}{P1}

Participants less familiar with AI and machine learning techniques found it particularly challenging to interpret the scores. This is because machine learning features may not have an explicit causal relationship with the predictions:

\quotes{I don't know anything about machine learning. I'm sure these variables connect in a way that I don't understand. But like, I don't understand why the number of authors makes a difference, or the citations, or where the authors are from. I think p-values are important, but I also think it depends on like, what were your stimuli? who were the participants? Was it a generalizable sample? You know, where did you get it from? Was it an experiment or survey? I feel like the methods play more of a role, whereas I feel like this is not looking at methods.}{P4}

P4 expressed concerns that features like the number of authors or citations may not be reliable indicators of replicability. Similarly, P14 questioned whether recently published papers would be unfairly deemed less reproducible due to a lower citation count or if the system would be ineffective for qualitative studies that don't rely on p-values. According to the participants, the trustworthiness of AI-generated results depends on whether specific features serve as reasonable predictors for replicability. As P8 suggested, some descriptive summary accompanying the estimation score could aid in interpreting the results:

\quotes{The conclusion is there, but it's very limited. If it could just say it's based on certain features that stood out, like, no citations, or the sample size, or whatever it's using. Not so much quantitative like in these numbers, but more of a narrative, that would be helpful.}{P8}

The lack of sufficient explainability led participants to hesitate to rely on AI for estimating replicability. As P17 noted, while the system could be useful for making predictions, using it for assessments would require stronger evidence, given that accusations of non-replicability carry significant weight:
\quotes{We need to have people redo the whole experiment (to decide reproducibility). The critics that others' work is not reproducible is a very strong accusation. So, you need to have very strong evidence to show that something is wrong.}{P17}

Moreover, an indecisive estimation score further eroded participants' confidence in the system. For instance, a score of around 50 made it hard for participants to form a strong opinion. Similarly, a bimodal distribution of results from all agents (Figure \ref{fig:results}) raised additional questions:

\quotes{It's a really bimodal distribution with a lot of bots thinking that it's not reproducible, and a lot think that it is reproducible. So I would wonder what the factors that contributed to that? Without knowing why it's bimodal, I would probably trust the results less.}{P11}

A bimodal distribution could signal high uncertainty, causing distrust. AI-estimated replicability was novel to most participants, and their trust might improve through continued interaction with the tool. P12 suggested that gaining confidence in the tool would require practice with multiple papers and analysis of result variability. Providing context for participants to compare and interpret scores could also enhance their trust:

\quotes{ I would like to have some reference scores. For example, in this research field, 80\% of the empirical articles are receiving this score, so that's why 59 is a good score for this field. Comparisons would make me feel more confident about the result here.}{P9}

In summary, although there was no outright rejection of AI-estimated replicability, participants were hesitant to trust the results fully. Providing detailed and transparent explanations of the AI system's workings and how each feature contributes to the estimation is essential for users to make informed decisions. Additionally, exposure to multiple papers or having reference points to interpret scores could further help users assess the tool's utility. 

\subsubsection{Potential use of AI-estimated replicability scores}
\label{potential}
While the prototype aims to offer researchers a level of confidence in published studies, participants expressed diverse opinions on its potential benefits. Firstly, some participants questioned the value of providing an estimated replicability score. As referenced in \ref{perception}, P15 argued that the concept is less relevant to qualitative and context-sensitive studies, doubting the system's usefulness. Similarly, P2 felt that the score's relevance was minimal as his priority in the literature search was the paper's relevance to his research. P6, on the other hand, considered the citation count as a sufficient indicator of paper quality:

\quotes{For the paper I just mentioned, I believe it has over 5000 citations. So that's definitely it's a good sign. This is publicly available and can be easily obtained. And if I already have this type of information, why bother using a more complicated system?}{P6}

Others were more optimistic about integrating the system into their research workflow. P10, for instance, felt that the estimated score could enrich the literature review process by flagging potential methodological flaws:

\quotes{I would imagine, as I was writing my literature review, I would be checking papers. And I would be saying, hey, in my literature review, I'm citing this paper, because it's foundational to the way we think about things. But there are concerns about whether it would be reproducible.}{P10}

In this scenario, the estimation could help identify weaknesses in foundational literature, thereby improving the rigor of literature reviews. Similarly, P9 envisioned the tool as a means to quantify the strength of research arguments based on cited references:

\quotes{If I have enough time, I would prepare a table with the empirical articles that I used and the scores that they have received to increase the credibility of my arguments. These empirical papers have high reproducible scores, which I use to support my arguments therefore my arguments are credible.}{P9}

Specifically, P9 saw the tool as useful for addressing reviewers' concerns about study designs during the peer-review process, citing high replicability scores as justification for chosen research methods.

Given that publishing is a primary goal for researchers, many participants considered the tool beneficial for peer review and similar contexts. Although originally designed for evaluating published work, P13 suggested using it as a pre-submission checker, particularly benefiting researchers with less publication experience:

\quotes{Especially like an early career researcher or grad student could pump their own papers in and then see what these synthetic agents think about my paper and (estimate) what other people are going to think about my paper.}{P13}

This suggestion was seconded by P11, who wanted to use the tool to assess his own work and see how reproducible and replicable they were perceived to be. By contrast, P12 provided a different perspective as paper reviewers, and suggested that the system could help reviewers who were asked to review papers that they have conceptual and theoretical interests in, but don't have methodological sophistication:

\quotes{So if I got asked to review a paper that involved machine learning, which was conceptually and theoretically involved in my area, and that's why I was chosen as a reviewer, I would probably use this tool as a kind of a cross-check against my thoughts on the paper.}{P11}

Similarly, P15 proposed that funding institutions could use the score in evaluating grant proposals:
\quotes{I think institutions such as the National Science Foundation, you know, they review grant proposals for funding, right? And replicability in itself is very important issue in science, regardless of hardcore science or social science. So institutions might be able to use this score to tell whether the proposed project has reproducibility.}{P15}

Despite participants' interests in augmenting the peer review process with the estimated reproducibility, the functionalities of the tool need to be enhanced and expanded for unpublished papers to fulfill the usages suggested above. While participants were generally interested in using the estimated reproducibility to augment the peer review process, the prototype would require further enhancements to cater to unpublished papers. Most participants agreed that more information would at least be interesting, if not particularly helpful. However, one participant warned against the unintended consequences of relying too heavily on algorithmically generated scores. Researchers might make efforts to boost the score of their papers without actually enhancing the research quality:

\quotes{I see it as helpful but also potentially harmful. Like if someone doesn't work out their paper in a certain way that is similar to other papers, but it's still reproducible? I can imagine there would be consequences for those papers that wrongfully hurt people's careers... With this competence score, I can see some strange reactions to it from scholars like trying to maximize the reproducibility score, where it's not really adding to the science, it's just adding to the perception of how people think of it.}{P3}

In summary, while the estimated replicability score might not necessarily facilitate the literature search or review process for all, it could serve as a valuable tool for supporting research arguments, checking paper quality, and aiding in peer review. It should be noted that all suggestions are made based on the assumption that the estimation is fully reliable, which requires significant research efforts to improve the current tool. More importantly, further investigations on ethical implications are essential for future iterations of AI-estimated replicability.

\section{Discussion}
\label{discussion}
Through qualitative interviews with 17 social science researchers, the current study explores design opportunities for technologies to support better literature review processes taking into consideration reproducibility and replicability. We present preliminary findings for three research questions. 1) We illustrate participants' approaches to literature search and review, highlighting the use of various indicators and utility-oriented information seeking (Section \ref{metrics} \& \ref{sec:evaluation}). 2) We identify several challenges participants encountered during the literature search, including the lack of expertise and limited capabilities of keywords-based search. 3) In terms of reproducibility, participants' understanding of the concept was blurred with perceptions of general research quality. Nonetheless, their feedback demonstrated the potential benefits of providing quantitative estimation of reproducibility for scientific publication. However, they also raised concerns about the system's explainability and interoperability and the ethical considerations of employing such a system. In the following sections, we contextualize these findings within relevant literature and discuss design implications as well as future research directions.

\subsection{Future technologies for supporting literature review}
Conducting effective literature search and review is foundational to research. However, the sheer volume of publications has made the task of identifying relevant literature increasingly challenging. While the HCI community has developed several design prototypes to potentially facilitate literature search and discovery, researchers predominantly rely on established bibliographic databases like Web of Science and Google Scholar. Consequently, understanding researchers' current experiences during literature search is crucial for developing effective and widely adopted technologies. Our findings about literature search align with prior research on user experiences with web search, highlighting the need for search tools that can handle more complex search queries \cite{white2018opportunities} and the impact of domain expertise in effective search \cite{mao2018does}. Yet, literature search brings a set of unique metrics which might inform, prioritize, and provide necessary context to search reseults. These findings also align with prior research emphasizing engineers' and technologists' use of publication venues and author names in search \cite{arshad2019scholarly} and confirm some of the challenges encountered by computer scientists, such as keeping up to date with research and exploring unfamiliar topics \cite{athukorala2013information}. These considerations can inform the design of better literature search tools, as discussed below.

\subsubsection{Design implications for literature search tools}
\textbf{Domain specific search interface:} Past research highlights Google Scholar's advantages, such as its user-friendly, easily navigable interface and extensive resource base \cite{martin-martin_evidence_2018, boeker_google_2013}. Our participants corroborated these points, stating a preference for Google Scholar. Although Google Scholar supports a smooth user experience, our participants expressed a need for more advanced features. One major limitation is that Google Scholar returns less specific and potentially lower-quality results compared to other databases. Some participants felt their expertise and familiarity with the topic helped mitigate this shortcoming. However, this presents an obstacle for novice researchers or those exploring an unfamiliar field, who may lack foundational knowledge about recognized venues and authors in that space. 

Venue reputation and author reputation are used as indicators for good papers. Although impact factors and h-indexes provide some insight, participants largely relied on their own experience to assess these reputations. Some design prototypes have indirectly aided users in identifying key researchers through citation network visualization \cite{sultanum_understanding_2020, choe_papers101_2021, dunne_rapid_2012}, but there is scant focus on venue and author reputation in existing literature search tool designs. 

These findings highlight the benefits of incorporating domain-specific expertise in information search. Domain expertise can be used to present better results and query suggestions to users \cite{white2009characterizing}. Therefore, the design of domain-specific search tools could be customized accordingly. For literature search tools, the user experience could be improved through better organization of search results based on expert knowledge: 1) enable users to filter results based on research areas; 2) recommend well-recognized publication venues; and 3) recommend well-recognized authors and allow users to filter results accordingly.

While these functionalities could help users find desired papers more efficiently, there are ethical considerations that should be further explored. Specifically, exclusive focus on well-known venues and authors could hinder the progress of science, which also relies on contributions from newcomers. Our participants have largely used citation count as an indicator of paper quality, with some mentioning the bias against newer publications, which naturally have fewer citations. Researchers have proposed alternative metrics such as weighted citation counts and Altmetrics \cite{bornmann_altmetrics_2014,yan_weighted_2010}. Future research on literature search should investigate how we can leverage these metrics to introduce emerging researchers while facilitating the discovery of foundational works.\\

\vspace{-0.3cm}
\noindent \textbf{Support complex user queries:} Literature search tools and algorithms should also be designed to handle more complex user queries. First, the quality of keyword-based search results is highly dependent on the keywords entered. \citet{choe_papers101_2021} observed that novice researchers often struggle with academic terminologies and proposed an interface that recommends relevant keywords based on the current search queries. Our study confirmed this challenge but also underscored the usefulness of providing keyword alternatives that capture synonymous research concepts, as this was a common struggle among participants.

More crucially, keyword-based search methods alone may be insufficient for fulfilling the nuanced needs of the search process. Participants expressed a desire not just to locate papers on specific research topics but also to find research that employs particular methodologies, utilizes specific datasets, or even adopts a particular viewpoint on a research issue. While the exact algorithm behind the Google Scholar search engine remains proprietary, it appears incapable of handling such complex queries. One possible solution is to leverage the fast-advancing capabilities of large language models (LLMs) \cite{birhane_science_2023,koneru2023can}. Preliminary work has already been done on AI-based tools that can identify pertinent papers and summarize their key findings based on user-generated questions rather than mere keywords \cite{kung_elicit_2023}. Our study suggests that this approach may better satisfy the needs of researchers. Therefore, future research should seek to evaluate the capacity of LLMs in handling complex search queries, and investigate the impact on user experience. In terms of technological design, the search tools should leverage LLMs to achieve the following: 1) identify and suggest alternative research terminologies that may increase the performance of user queries; and 2) perform deep analysis of literature to answer specific user queries on research questions, methods, and findings. These functionalities may accelerate the process of literature review, helping scholars to identify and organize papers that meet specific criteria without going through a long list of search results. Scholars can be empowered to understand theories, methodological approaches, and certain hypotheses more efficiently, which facilitates the process of scientific discovery. While promising, research is also needed to evaluate the verifiability of results from search engines empowered by LLMs \cite{liu2023evaluating}.

\subsubsection{Problems with scientific paper reading}
After identifying relevant papers, researchers also invest considerable time in reading and understanding them. Prior studies have developed tools that aim to facilitate the reading process, using techniques such as tooltips and Natural Language Processing (NLP) to augment the text summarization\cite{head_augmenting_2021,chang_citesee_2023,palani_relatedly_2023}. While our study did not directly identify challenges in the reading process, it did reveal significant variability in participants' reading strategies. Specifically, they prioritized different paper sections based on their information-seeking needs.

This observation raises important questions. First, should this utility-oriented reading style be encouraged? Extant literature offers varied perspectives on effective approaches to reading and understanding a scientific paper \cite{keshav_how_2007,greenhalgh_how_1997}. Some suggest deviating from the standard IMRaD (Introduction, Methods, Results, and Discussion) structure for a more efficient understanding \cite{durbin_how_2009}. While reading papers out of order or skipping sections may increase efficiency, there is a risk of neglecting essential information, potentially leading to misunderstandings and inappropriate referencing. The second question is more complicated: is it time to reconsider the IMRaD structure? The structure of scientific papers has been guided by long-standing traditions and norms, and despite criticisms regarding its rigidity \cite{sollaci_introduction_2004}, the research community continues to adhere to it. In light of the changes brought about by computer-mediated communication, some researchers advocate for new forms of scholarly communication \cite{ponte_scholarly_2011,dashnow_ten_2014}. This exploratory study provides preliminary insights into this issue. Instead of concrete design implications, we believe that more thorough examinations of the following are required for supporting scientific paper reading: 1) how augmentative technologies can help readers locate information in a scientific report; 2) whether the use of such technologies would lead to lack-of-context and biased interpretations of a scientific report; and eventually 3) whether we need more modern and diverse formats for scientific reporting. The extension of reporting beyond IMRaD can accommodate the diversity of research methodologies and genres \cite{eriksson2023art} and potentially facilitate the identification and reading of literature. In addition, the investigation of the misalignment between IMRaD and reading behaviors will inform teaching in scientific writing and help scholars to communicate their work in a manner that effectively captures attention from broad audiences \cite{shiely2024and}.

\subsection{Promises and risks of replicability estimation}
Replicability and reproducibility are critical to the integrity of scientific progress. The replicability crisis has gained widespread recognition across disciplines, leading the scientific community to propose various initatives aimed at fostering reproducible research practices \cite{foster_open_2017,de_chaumont_icy_2012}. This study contributes a unique perspective by examining the implications of quantitative metrics on the literature search and review process.

\subsubsection{Promote understanding of key concepts}
While participants in this study were generally aware of the concepts of reproducibility and reproducibility and acknowledged their importance, their definitions of these terms varied. This discrepancy underscores the potential need to address these concepts more rigorously in graduate education. For instance, P7's statement about p-values showed a common misconception about p-values \cite{wasserstein2016asa} and, therefore, an incorrect definition of replication. Some participants confused the concepts of generalizability with reproducibility. This finding suggests that many researchers may have a nebulous understanding of replicability as a concept generally related to research quality. This general understanding does not undermine the validity of the study as we seek to reflect on participants' current understanding and daily practices. However, it 
also suggests that research is needed to help design curriculum and educational resources that promote researchers' understanding of key concepts, which can hardly be achieved through the prediction tool.

Participants P3 and P9 emphasized that specific characteristics of human behavior studies, such as qualitative interviews, are challenging to replicate, calling into question the applicability of replicability to these types of studies. Heightened awareness of the replicability crisis has primarily arisen from publications focusing on quantitative and statistical methods \cite{baker_1500_2016}. However, recent arguments from qualitative researchers suggest that even in studies not aimed at hypothesis testing, preregistration practices can enhance reproducibility by serving as a check on subjectivity \cite{l_haven_preregistering_2019}. The reproducibility of qualitative research remains a significantly underexplored area \cite{cole_integrative_2023}. Our findings indicate a need to include more qualitative researchers in discussions about reproducibility to broaden awareness and advocate for reproducible practices in qualitative studies.

\textbf{Design opportunities for literature review:} Contrary to the importance of replicability in current scientific dialogue, participants did not explicitly seek out indicators of replicability or reproducibility during their literature searches. Instead, they implicitly assessed these factors by scrutinizing the methodologies described in the papers they read. While such assessments can be somewhat subjective, objective indicators like data availability, code availability, and preregistration practices could bolster readers' confidence in a study. As such, literature search tool designers should consider highlighting these indicators in search interfaces and enable users to prioritize studies that adhere to reproducible practices. Previous studies have shown that open-access (OA) articles are cited slightly more frequently than non-OA articles \cite{tamminen_open_2018}. Still, there is a noticeable gap in research exploring the impact of reproducible practices on citation rates. This study presents preliminary evidence that users pay attention to credibility markers when evaluating scholarly work. Future studies should examine how incorporating these credibility markers into the literature search experience may encourage researchers to prioritize publications adhering to recommended research practices.

\subsubsection{Transparent AI estimation of replicability}
\label{transparent}
In this study, we collected user feedback on a prototype AI to estimate replicability, building upon prior work \cite{rajtmajer2021synthetic}. Rather than assessing the accuracy of the prototype's estimations, our focus lied in understanding how a quantitative metric for reproducibility and reproducibility might impact the literature review process. Broadly speaking, participants expressed hesitancy to trust and use the AI-generated estimations, attributing their skepticism mainly to the interface's lack of explainability. Explainable AI (XAI) is crucial for facilitating user understanding in AI applications \cite{samek_towards_2019}. The current tool provides limited explanations of the decision-making process and related uncertainty. Specifically, they were unclear about which features most influenced a paper's estimated replicability, requesting quantification of the impact of each input on the output. This is a common technique used for developing more transparent systems \cite{linardatos_explainable_2021}. More critically, participants also expressed concern over some of the metrics used by the tool, such as the number of authors and university ranking, which they considered counter-intuitive. Generally, researchers tend to rely on methodological rigor to assess reproducibility, rather than these auxiliary metrics. Machine learning models tend to capture correlations in data to make predictions; however, researchers advocate that XAI should aim to identify causal relationships to better align with human understanding \cite{chou_counterfactuals_2022}. This issue is especially important for replicability estimations. The decision of whether a publication is replicable often comes with significant ethical implications for both authors and readers. Our results suggest that the major reliance on non-causal relationships in replicability estimations may not be acceptable for researchers. Future research should try to address this issue from two aspects: 1) improve the transparency of model decision-making and enhance the presentation of decision uncertainty, which can facilitate human-AI collaboration by motivating users to think more vigilantly about the results; 2) seek features that can represent the actual causal relationship with replicability, striking a balance between model efficiency and socio-technical benefits.

\textbf{Design implications for quantifying replicability:} One important improvement suggested by participants is the contextualization of estimation scores. On the one hand, the state of reproducibility can vary significantly across disciplines, rendering a 0-100 range too general for universal application. On the other hand, a score like 50 can appear ambiguous to users, necessitating additional context for proper interpretation. As \citet{gunning_xaiexplainable_2019} have posited, the definitions of interpretability and explainability in XAI may be domain-specific. Our study, as an initial exploration, suggests that within the realm of reproducibility, enhancing both transparency in decision-making and contextualization of results is critical for users to effectively interpret AI and decide whether and how they will use the output. In the design of AI and machine learning systems, explainability and interpretability should be emphasized in order to obtain improvement feedback from users and promote better human-AI collaboration.

\subsubsection{Benefits and risks of estimating replicability}

Replicability estimation is more than just providing a score to users. The wide applications of such systems may have a mixed impact on research activities and important ethical considerations. On the bright side, participants speculated that the score could serve as a subjective gauge of integrity for their literature review, support their methodological choices, or even supplement traditional peer review processes. This suggests that participants view reproducibility and replicability as important criteria for research quality and that a quantitative metric could facilitate their evaluation of scholarly works. While previous studies have associated research quality with reproducibility, they also highlight a lack of uniformity in evaluating reproducibility \cite{heroux_quality_2022}. Our findings suggest that providing a quantitative metric could help researchers identify replicable work to serve as foundational premises for their own research. Therefore, introducing the replicability estimation tool can positively impact the research community, promoting awareness of and emphasis on replicable and reproducible work. Researchers could easily determine if findings from a prior publication can stand the test of scrutiny, which can be incorporated into literature review, peer review, and publication process. As the current research culture often disproportionately favors novel and affirmative results \cite{munafo_manifesto_2017}, such an impact aids in upholding the integrity of scientific progress but also fosters a more transparent and trustworthy research environment. However, it should noted that these potential benefits are based on the assumption that the metric is fully reliable and trustworthy. As discussed in Section \ref{transparent}, there are several issues to be addressed before the AI estimation can be accepted as a valid indicator for reproducibility.

More importantly, there are potential risks associated with quantifying replicability, as participants have warned against during the interview. Specifically, they were concerned that researchers might chase higher scores in ways that could be counterproductive or unethical. Indeed, it has been criticized by the research community that efforts to enhance reproducibility may not necessarily promote high-quality research \cite{leonelli_rethinking_2018} and the over-emphasize of reproducibility may hinder the progress of scientific discoveries to some extent \cite{arshad2019scholarly}. Another fundamental issue with this tool is the validity and reliability of its output, as its capability to handle the numerous, ever-increasing future publications remains uncertain. Developing safe and useful AI systems will require us to make progress on scalable oversight \cite{bowman2022measuring}, which enables continuous supervision of AI performance. As such, further investigations are required to explore the practical and ethical ramifications of quantifying replicability, especially concerning how it may alter research practices. The current study serves as a starting point for examining replicability metrics. The preliminary findings suggest that providing indicators for replicability or research quality, in general, could potentially benefit researchers by supporting literature review, yet the design of such metrics must carefully consider transparency and explainability of the interface as well as the ethical implications of deploying the system. In addition to technical improvements, policy, and institutional initiatives are also desirable to ensure the quantification of replicability will lead to advancements in the quality and integrity of scientific research.

\section{Limitations and future work}
This study has several limitations that future research should address. First, as we have discussed, even though open code and open data do not typically apply to qualitative studies, they can still benefit from pre-registration. The current study is not pre-registered due to its exploratory nature and lack of specific hypotheses. This is a limitation that will be addressed in future confirmatory studies. Second, we relied on a prototype built upon existing work. Although it offers advanced functionalities, there is scope for further refinement, and alternative designs for AI-based replicability estimation could present different advantages and drawbacks. Third, our study used a lab-based demonstration approach to gather feedback. Conducting a longitudinal study that involves the tool's usage in real research settings would provide a more in-depth understanding of its impact on the research life cycle.

Several pertinent questions arose from our findings, although they fall outside the scope of this study. For instance, participants exhibited a somewhat nebulous understanding of reproducibility and replicability, which was further complicated by other concerns related to research validity. While these concerns are intrinsically linked to research quality, the consequences of this overlapping understanding merit further investigation. Furthermore, participants questioned the rationale behind using metrics such as citation counts and author reputation to estimate replicability, despite admitting to using these very metrics in their own literature searches. Future research should explore to what extent this reliance on traditional metrics is a byproduct of community culture rather than an indicator of research quality.

\section{Conclusion}
This study adopted a qualitative approach to explore potential enhancements to the literature search, particularly through the integration of reproducibility and replicability estimation. By examining participants' existing literature search practices, we identified key metrics that influence search result filtering. These metrics—such as venue reputation, author reputation, citation counts, and publication recency—are only partially supported by current search tools. There is a clear need for more intelligent search tools that offer users flexibility in retrieving information that suits their requirements at various stages of research. While incorporating replicability metrics into search processes has the potential to support researchers, we observed mixed attitudes towards AI-driven replicability estimations. Issues related to explainability, interpretability, and causality within AI models should be addressed before users can assess the usefulness of such systems. Assuming that these estimations are trustworthy, they could be used to strengthen research design and even facilitate the peer-review process. However, this comes with ethical implications and an undue emphasis on reproducibility may not necessarily result in better research quality, demanding further investigations. Overall, this study offers valuable insights for enhancing the design of literature search and management tools. It also underscores the need for further reseach and discussion on integrating replicability metrics, addressing ethical considerations, and exploring the balance between reproducibility and research quality.

\bibliographystyle{ACM-Reference-Format}
\bibliography{main}


\begin{thebibliography}{119}


\ifx \showCODEN    \undefined \def \showCODEN     #1{\unskip}     \fi
\ifx \showDOI      \undefined \def \showDOI       #1{#1}\fi
\ifx \showISBNx    \undefined \def \showISBNx     #1{\unskip}     \fi
\ifx \showISBNxiii \undefined \def \showISBNxiii  #1{\unskip}     \fi
\ifx \showISSN     \undefined \def \showISSN      #1{\unskip}     \fi
\ifx \showLCCN     \undefined \def \showLCCN      #1{\unskip}     \fi
\ifx \shownote     \undefined \def \shownote      #1{#1}          \fi
\ifx \showarticletitle \undefined \def \showarticletitle #1{#1}   \fi
\ifx \showURL      \undefined \def \showURL       {\relax}        \fi
\providecommand\bibfield[2]{#2}
\providecommand\bibinfo[2]{#2}
\providecommand\natexlab[1]{#1}
\providecommand\showeprint[2][]{arXiv:#2}

\bibitem[Al-Zubidy and Carver(2019)]%
        {al2019identification}
\bibfield{author}{\bibinfo{person}{Ahmed Al-Zubidy} {and}
  \bibinfo{person}{Jeffrey~C Carver}.} \bibinfo{year}{2019}\natexlab{}.
\newblock \showarticletitle{Identification and prioritization of SLR search
  tool requirements: an SLR and a survey}.
\newblock \bibinfo{journal}{\emph{Empirical Software Engineering}}
  \bibinfo{volume}{24} (\bibinfo{year}{2019}), \bibinfo{pages}{139--169}.
\newblock


\bibitem[Ali et~al\mbox{.}(2023)]%
        {ali2023explainable}
\bibfield{author}{\bibinfo{person}{Sajid Ali}, \bibinfo{person}{Tamer Abuhmed},
  \bibinfo{person}{Shaker El-Sappagh}, \bibinfo{person}{Khan Muhammad},
  \bibinfo{person}{Jose~M Alonso-Moral}, \bibinfo{person}{Roberto
  Confalonieri}, \bibinfo{person}{Riccardo Guidotti}, \bibinfo{person}{Javier
  Del~Ser}, \bibinfo{person}{Natalia D{\'\i}az-Rodr{\'\i}guez}, {and}
  \bibinfo{person}{Francisco Herrera}.} \bibinfo{year}{2023}\natexlab{}.
\newblock \showarticletitle{Explainable Artificial Intelligence (XAI): What we
  know and what is left to attain Trustworthy Artificial Intelligence}.
\newblock \bibinfo{journal}{\emph{Information Fusion}}  \bibinfo{volume}{99}
  (\bibinfo{year}{2023}), \bibinfo{pages}{101805}.
\newblock


\bibitem[Altmejd et~al\mbox{.}(2019)]%
        {altmejd2019predicting}
\bibfield{author}{\bibinfo{person}{Adam Altmejd}, \bibinfo{person}{Anna
  Dreber}, \bibinfo{person}{Eskil Forsell}, \bibinfo{person}{Juergen Huber},
  \bibinfo{person}{Taisuke Imai}, \bibinfo{person}{Magnus Johannesson},
  \bibinfo{person}{Michael Kirchler}, \bibinfo{person}{Gideon Nave}, {and}
  \bibinfo{person}{Colin Camerer}.} \bibinfo{year}{2019}\natexlab{}.
\newblock \showarticletitle{Predicting the replicability of social science lab
  experiments}.
\newblock \bibinfo{journal}{\emph{PloS one}} \bibinfo{volume}{14},
  \bibinfo{number}{12} (\bibinfo{year}{2019}), \bibinfo{pages}{e0225826}.
\newblock


\bibitem[Ammari and Schoenebeck(2015)]%
        {ammari2015understanding}
\bibfield{author}{\bibinfo{person}{Tawfiq Ammari} {and} \bibinfo{person}{Sarita
  Schoenebeck}.} \bibinfo{year}{2015}\natexlab{}.
\newblock \showarticletitle{Understanding and supporting fathers and fatherhood
  on social media sites}. In \bibinfo{booktitle}{\emph{Proceedings of the 33rd
  annual ACM conference on human factors in computing systems}}.
  \bibinfo{pages}{1905--1914}.
\newblock


\bibitem[Arshad and Ameen(2019)]%
        {arshad2019scholarly}
\bibfield{author}{\bibinfo{person}{Alia Arshad} {and} \bibinfo{person}{Kanwal
  Ameen}.} \bibinfo{year}{2019}\natexlab{}.
\newblock \showarticletitle{Scholarly information seeking of academic engineers
  and technologists}.
\newblock \bibinfo{journal}{\emph{International Information \& Library Review}}
  \bibinfo{volume}{51}, \bibinfo{number}{1} (\bibinfo{year}{2019}),
  \bibinfo{pages}{1--8}.
\newblock


\bibitem[Athukorala et~al\mbox{.}(2016)]%
        {athukorala2016exploratory}
\bibfield{author}{\bibinfo{person}{Kumaripaba Athukorala},
  \bibinfo{person}{Dorota G{\l}owacka}, \bibinfo{person}{Giulio Jacucci},
  \bibinfo{person}{Antti Oulasvirta}, {and} \bibinfo{person}{Jilles Vreeken}.}
  \bibinfo{year}{2016}\natexlab{}.
\newblock \showarticletitle{Is exploratory search different? A comparison of
  information search behavior for exploratory and lookup tasks}.
\newblock \bibinfo{journal}{\emph{Journal of the Association for Information
  Science and Technology}} \bibinfo{volume}{67}, \bibinfo{number}{11}
  (\bibinfo{year}{2016}), \bibinfo{pages}{2635--2651}.
\newblock


\bibitem[Athukorala et~al\mbox{.}(2013)]%
        {athukorala2013information}
\bibfield{author}{\bibinfo{person}{Kumaripaba Athukorala}, \bibinfo{person}{Eve
  Hoggan}, \bibinfo{person}{Anu Lehti{\"o}}, \bibinfo{person}{Tuukka Ruotsalo},
  {and} \bibinfo{person}{Giulio Jacucci}.} \bibinfo{year}{2013}\natexlab{}.
\newblock \showarticletitle{Information-seeking behaviors of computer
  scientists: Challenges for electronic literature search tools}.
\newblock \bibinfo{journal}{\emph{Proceedings of the American Society for
  Information Science and Technology}} \bibinfo{volume}{50},
  \bibinfo{number}{1} (\bibinfo{year}{2013}), \bibinfo{pages}{1--11}.
\newblock


\bibitem[Atkinson and Cipriani(2018)]%
        {atkinson_how_2018}
\bibfield{author}{\bibinfo{person}{Lauren~Z. Atkinson} {and}
  \bibinfo{person}{Andrea Cipriani}.} \bibinfo{year}{2018}\natexlab{}.
\newblock \showarticletitle{How to carry out a literature search for a
  systematic review: a practical guide}.
\newblock \bibinfo{journal}{\emph{BJPsych Advances}} \bibinfo{volume}{24},
  \bibinfo{number}{2} (\bibinfo{date}{March} \bibinfo{year}{2018}),
  \bibinfo{pages}{74--82}.
\newblock
\showISSN{2056-4678, 2056-4686}
\urldef\tempurl%
\url{https://doi.org/10.1192/bja.2017.3}
\showDOI{\tempurl}


\bibitem[Baker(2016a)]%
        {baker_1500_2016}
\bibfield{author}{\bibinfo{person}{Monya Baker}.}
  \bibinfo{year}{2016}\natexlab{a}.
\newblock \showarticletitle{1,500 scientists lift the lid on reproducibility}.
\newblock \bibinfo{journal}{\emph{Nature}} \bibinfo{volume}{533},
  \bibinfo{number}{7604} (\bibinfo{date}{May} \bibinfo{year}{2016}),
  \bibinfo{pages}{452--454}.
\newblock
\showISSN{1476-4687}
\urldef\tempurl%
\url{https://doi.org/10.1038/533452a}
\showDOI{\tempurl}
\newblock
\shownote{Number: 7604 Publisher: Nature Publishing Group}.


\bibitem[Baker(2016b)]%
        {baker_reproducibility_2016}
\bibfield{author}{\bibinfo{person}{Monya Baker}.}
  \bibinfo{year}{2016}\natexlab{b}.
\newblock \showarticletitle{Reproducibility crisis}.
\newblock \bibinfo{journal}{\emph{Nature}} \bibinfo{volume}{533},
  \bibinfo{number}{26} (\bibinfo{year}{2016}), \bibinfo{pages}{353--66}.
\newblock


\bibitem[Baker(2000)]%
        {baker_writing_2000}
\bibfield{author}{\bibinfo{person}{Michael~J. Baker}.}
  \bibinfo{year}{2000}\natexlab{}.
\newblock \showarticletitle{Writing a literature review}.
\newblock \bibinfo{journal}{\emph{The marketing review}} \bibinfo{volume}{1},
  \bibinfo{number}{2} (\bibinfo{year}{2000}), \bibinfo{pages}{219--247}.
\newblock
\newblock
\shownote{ISBN: 1469-347X Publisher: Westburn Publishers Ltd}.


\bibitem[Bartholomew(2014)]%
        {bartholomew2014science}
\bibfield{author}{\bibinfo{person}{Robert~E Bartholomew}.}
  \bibinfo{year}{2014}\natexlab{}.
\newblock \bibinfo{title}{Science for sale: the rise of predatory journals}.
\newblock , \bibinfo{numpages}{384--385}~pages.
\newblock


\bibitem[Berry et~al\mbox{.}(2017)]%
        {berry2017assessing}
\bibfield{author}{\bibinfo{person}{James Berry}, \bibinfo{person}{Lucas~C
  Coffman}, \bibinfo{person}{Douglas Hanley}, \bibinfo{person}{Rania Gihleb},
  {and} \bibinfo{person}{Alistair~J Wilson}.} \bibinfo{year}{2017}\natexlab{}.
\newblock \showarticletitle{Assessing the rate of replication in economics}.
\newblock \bibinfo{journal}{\emph{American Economic Review}}
  \bibinfo{volume}{107}, \bibinfo{number}{5} (\bibinfo{year}{2017}),
  \bibinfo{pages}{27--31}.
\newblock


\bibitem[Bethard and Jurafsky(2010)]%
        {bethard_who_2010}
\bibfield{author}{\bibinfo{person}{Steven Bethard} {and} \bibinfo{person}{Dan
  Jurafsky}.} \bibinfo{year}{2010}\natexlab{}.
\newblock \showarticletitle{Who should {I} cite: learning literature search
  models from citation behavior}. In \bibinfo{booktitle}{\emph{Proceedings of
  the 19th {ACM} international conference on {Information} and knowledge
  management}}. \bibinfo{publisher}{ACM}, \bibinfo{address}{Toronto ON Canada},
  \bibinfo{pages}{609--618}.
\newblock
\showISBNx{978-1-4503-0099-5}
\urldef\tempurl%
\url{https://doi.org/10.1145/1871437.1871517}
\showDOI{\tempurl}


\bibitem[Birhane et~al\mbox{.}(2023)]%
        {birhane_science_2023}
\bibfield{author}{\bibinfo{person}{Abeba Birhane}, \bibinfo{person}{Atoosa
  Kasirzadeh}, \bibinfo{person}{David Leslie}, {and} \bibinfo{person}{Sandra
  Wachter}.} \bibinfo{year}{2023}\natexlab{}.
\newblock \showarticletitle{Science in the age of large language models}.
\newblock \bibinfo{journal}{\emph{Nature Reviews Physics}}
  (\bibinfo{year}{2023}), \bibinfo{pages}{1--4}.
\newblock
\newblock
\shownote{ISBN: 2522-5820 Publisher: Nature Publishing Group UK London}.


\bibitem[Boeker et~al\mbox{.}(2013)]%
        {boeker_google_2013}
\bibfield{author}{\bibinfo{person}{Martin Boeker}, \bibinfo{person}{Werner
  Vach}, {and} \bibinfo{person}{Edith Motschall}.}
  \bibinfo{year}{2013}\natexlab{}.
\newblock \showarticletitle{Google {Scholar} as replacement for systematic
  literature searches: good relative recall and precision are not enough}.
\newblock \bibinfo{journal}{\emph{BMC medical research methodology}}
  \bibinfo{volume}{13}, \bibinfo{number}{1} (\bibinfo{year}{2013}),
  \bibinfo{pages}{1--12}.
\newblock
\newblock
\shownote{ISBN: 1471-2288 Publisher: BioMed Central}.


\bibitem[Bornmann(2014)]%
        {bornmann_altmetrics_2014}
\bibfield{author}{\bibinfo{person}{Lutz Bornmann}.}
  \bibinfo{year}{2014}\natexlab{}.
\newblock \showarticletitle{Do altmetrics point to the broader impact of
  research? {An} overview of benefits and disadvantages of altmetrics}.
\newblock \bibinfo{journal}{\emph{Journal of informetrics}}
  \bibinfo{volume}{8}, \bibinfo{number}{4} (\bibinfo{year}{2014}),
  \bibinfo{pages}{895--903}.
\newblock
\newblock
\shownote{ISBN: 1751-1577 Publisher: Elsevier}.


\bibitem[Botvinik-Nezer et~al\mbox{.}(2020)]%
        {botvinik2020variability}
\bibfield{author}{\bibinfo{person}{Rotem Botvinik-Nezer},
  \bibinfo{person}{Felix Holzmeister}, \bibinfo{person}{Colin~F Camerer},
  \bibinfo{person}{Anna Dreber}, \bibinfo{person}{Juergen Huber},
  \bibinfo{person}{Magnus Johannesson}, \bibinfo{person}{Michael Kirchler},
  \bibinfo{person}{Roni Iwanir}, \bibinfo{person}{Jeanette~A Mumford},
  \bibinfo{person}{R~Alison Adcock}, {et~al\mbox{.}}}
  \bibinfo{year}{2020}\natexlab{}.
\newblock \showarticletitle{Variability in the analysis of a single
  neuroimaging dataset by many teams}.
\newblock \bibinfo{journal}{\emph{Nature}} \bibinfo{volume}{582},
  \bibinfo{number}{7810} (\bibinfo{year}{2020}), \bibinfo{pages}{84--88}.
\newblock


\bibitem[Bowman et~al\mbox{.}(2022)]%
        {bowman2022measuring}
\bibfield{author}{\bibinfo{person}{Samuel~R Bowman}, \bibinfo{person}{Jeeyoon
  Hyun}, \bibinfo{person}{Ethan Perez}, \bibinfo{person}{Edwin Chen},
  \bibinfo{person}{Craig Pettit}, \bibinfo{person}{Scott Heiner},
  \bibinfo{person}{Kamil{\.e} Luko{\v{s}}i{\=u}t{\.e}}, \bibinfo{person}{Amanda
  Askell}, \bibinfo{person}{Andy Jones}, \bibinfo{person}{Anna Chen},
  {et~al\mbox{.}}} \bibinfo{year}{2022}\natexlab{}.
\newblock \showarticletitle{Measuring progress on scalable oversight for large
  language models}.
\newblock \bibinfo{journal}{\emph{arXiv preprint arXiv:2211.03540}}
  (\bibinfo{year}{2022}).
\newblock


\bibitem[Bramer et~al\mbox{.}(2018)]%
        {bramer_systematic_2018}
\bibfield{author}{\bibinfo{person}{Wichor~M. Bramer},
  \bibinfo{person}{Gerdien~B. de Jonge}, \bibinfo{person}{Melissa~L.
  Rethlefsen}, \bibinfo{person}{Frans Mast}, {and} \bibinfo{person}{Jos
  Kleijnen}.} \bibinfo{year}{2018}\natexlab{}.
\newblock \showarticletitle{A systematic approach to searching: an efficient
  and complete method to develop literature searches}.
\newblock \bibinfo{journal}{\emph{Journal of the Medical Library Association :
  JMLA}} \bibinfo{volume}{106}, \bibinfo{number}{4} (\bibinfo{date}{Oct.}
  \bibinfo{year}{2018}), \bibinfo{pages}{531--541}.
\newblock
\showISSN{1536-5050}
\urldef\tempurl%
\url{https://doi.org/10.5195/jmla.2018.283}
\showDOI{\tempurl}


\bibitem[Braun and Clarke(2006)]%
        {braun2006using}
\bibfield{author}{\bibinfo{person}{Virginia Braun} {and}
  \bibinfo{person}{Victoria Clarke}.} \bibinfo{year}{2006}\natexlab{}.
\newblock \showarticletitle{Using thematic analysis in psychology}.
\newblock \bibinfo{journal}{\emph{Qualitative research in psychology}}
  \bibinfo{volume}{3}, \bibinfo{number}{2} (\bibinfo{year}{2006}),
  \bibinfo{pages}{77--101}.
\newblock


\bibitem[Brocke et~al\mbox{.}(2009)]%
        {brocke_reconstructing_2009}
\bibfield{author}{\bibinfo{person}{Jan~vom Brocke}, \bibinfo{person}{Alexander
  Simons}, \bibinfo{person}{Bjoern Niehaves}, \bibinfo{person}{Bjorn Niehaves},
  \bibinfo{person}{Kai Reimer}, \bibinfo{person}{Ralf Plattfaut}, {and}
  \bibinfo{person}{Anne Cleven}.} \bibinfo{year}{2009}\natexlab{}.
\newblock \showarticletitle{Reconstructing the giant: {On} the importance of
  rigour in documenting the literature search process}.
\newblock  (\bibinfo{year}{2009}).
\newblock


\bibitem[Camerer et~al\mbox{.}(2016a)]%
        {camerer_evaluating_2016}
\bibfield{author}{\bibinfo{person}{Colin~F. Camerer}, \bibinfo{person}{Anna
  Dreber}, \bibinfo{person}{Eskil Forsell}, \bibinfo{person}{Teck-Hua Ho},
  \bibinfo{person}{Jürgen Huber}, \bibinfo{person}{Magnus Johannesson},
  \bibinfo{person}{Michael Kirchler}, \bibinfo{person}{Johan Almenberg},
  \bibinfo{person}{Adam Altmejd}, {and} \bibinfo{person}{Taizan Chan}.}
  \bibinfo{year}{2016}\natexlab{a}.
\newblock \showarticletitle{Evaluating replicability of laboratory experiments
  in economics}.
\newblock \bibinfo{journal}{\emph{Science}} \bibinfo{volume}{351},
  \bibinfo{number}{6280} (\bibinfo{year}{2016}), \bibinfo{pages}{1433--1436}.
\newblock
\newblock
\shownote{ISBN: 0036-8075 Publisher: American Association for the Advancement
  of Science}.


\bibitem[Camerer et~al\mbox{.}(2016b)]%
        {camerer2016evaluating}
\bibfield{author}{\bibinfo{person}{Colin~F Camerer}, \bibinfo{person}{Anna
  Dreber}, \bibinfo{person}{Eskil Forsell}, \bibinfo{person}{Teck-Hua Ho},
  \bibinfo{person}{J{\"u}rgen Huber}, \bibinfo{person}{Magnus Johannesson},
  \bibinfo{person}{Michael Kirchler}, \bibinfo{person}{Johan Almenberg},
  \bibinfo{person}{Adam Altmejd}, \bibinfo{person}{Taizan Chan},
  {et~al\mbox{.}}} \bibinfo{year}{2016}\natexlab{b}.
\newblock \showarticletitle{Evaluating replicability of laboratory experiments
  in economics}.
\newblock \bibinfo{journal}{\emph{Science}} \bibinfo{volume}{351},
  \bibinfo{number}{6280} (\bibinfo{year}{2016}), \bibinfo{pages}{1433--1436}.
\newblock


\bibitem[Camerer et~al\mbox{.}(2018)]%
        {camerer2018evaluating}
\bibfield{author}{\bibinfo{person}{Colin~F Camerer}, \bibinfo{person}{Anna
  Dreber}, \bibinfo{person}{Felix Holzmeister}, \bibinfo{person}{Teck-Hua Ho},
  \bibinfo{person}{J{\"u}rgen Huber}, \bibinfo{person}{Magnus Johannesson},
  \bibinfo{person}{Michael Kirchler}, \bibinfo{person}{Gideon Nave},
  \bibinfo{person}{Brian~A Nosek}, \bibinfo{person}{Thomas Pfeiffer},
  {et~al\mbox{.}}} \bibinfo{year}{2018}\natexlab{}.
\newblock \showarticletitle{Evaluating the replicability of social science
  experiments in Nature and Science between 2010 and 2015}.
\newblock \bibinfo{journal}{\emph{Nature human behaviour}} \bibinfo{volume}{2},
  \bibinfo{number}{9} (\bibinfo{year}{2018}), \bibinfo{pages}{637--644}.
\newblock


\bibitem[Cassenti and Kaplan(2021)]%
        {cassenti2021robust}
\bibfield{author}{\bibinfo{person}{Daniel~N Cassenti} {and}
  \bibinfo{person}{Lance~M Kaplan}.} \bibinfo{year}{2021}\natexlab{}.
\newblock \showarticletitle{Robust uncertainty representation in human-AI
  collaboration}. In \bibinfo{booktitle}{\emph{Artificial Intelligence and
  Machine Learning for Multi-Domain Operations Applications III}},
  Vol.~\bibinfo{volume}{11746}. SPIE, \bibinfo{pages}{249--262}.
\newblock


\bibitem[Cestero et~al\mbox{.}(2022)]%
        {cestero2022pysurveillance}
\bibfield{author}{\bibinfo{person}{Julen Cestero}, \bibinfo{person}{David
  Vel{\'a}squez}, \bibinfo{person}{Elizabeth Suesc{\'u}n},
  \bibinfo{person}{Mikel Maiza}, {and} \bibinfo{person}{Marco Quartulli}.}
  \bibinfo{year}{2022}\natexlab{}.
\newblock \showarticletitle{Pysurveillance: A Novel Tool for Supporting
  Researchers in the Systematic Literature Review Process}.
\newblock \bibinfo{journal}{\emph{Advanced Intelligent Technologies for
  Industry}} (\bibinfo{year}{2022}), \bibinfo{pages}{239--248}.
\newblock


\bibitem[Chang et~al\mbox{.}(2023)]%
        {chang_citesee_2023}
\bibfield{author}{\bibinfo{person}{Joseph~Chee Chang}, \bibinfo{person}{Amy~X.
  Zhang}, \bibinfo{person}{Jonathan Bragg}, \bibinfo{person}{Andrew Head},
  \bibinfo{person}{Kyle Lo}, \bibinfo{person}{Doug Downey}, {and}
  \bibinfo{person}{Daniel~S. Weld}.} \bibinfo{year}{2023}\natexlab{}.
\newblock \showarticletitle{{CiteSee}: {Augmenting} {Citations} in {Scientific}
  {Papers} with {Persistent} and {Personalized} {Historical} {Context}}. In
  \bibinfo{booktitle}{\emph{Proceedings of the 2023 {CHI} {Conference} on
  {Human} {Factors} in {Computing} {Systems}}}. \bibinfo{publisher}{ACM},
  \bibinfo{address}{Hamburg Germany}, \bibinfo{pages}{1--15}.
\newblock
\showISBNx{978-1-4503-9421-5}
\urldef\tempurl%
\url{https://doi.org/10.1145/3544548.3580847}
\showDOI{\tempurl}


\bibitem[Chen and Eickhoff(2022)]%
        {chen2022evaluating}
\bibfield{author}{\bibinfo{person}{Catherine Chen} {and}
  \bibinfo{person}{Carsten Eickhoff}.} \bibinfo{year}{2022}\natexlab{}.
\newblock \showarticletitle{Evaluating Search Explainability with Psychometrics
  and Crowdsourcing}.
\newblock \bibinfo{journal}{\emph{arXiv preprint arXiv:2210.09430}}
  (\bibinfo{year}{2022}).
\newblock


\bibitem[Choe et~al\mbox{.}(2021)]%
        {choe_papers101_2021}
\bibfield{author}{\bibinfo{person}{Kiroong Choe}, \bibinfo{person}{Seokweon
  Jung}, \bibinfo{person}{Seokhyeon Park}, \bibinfo{person}{Hwajung Hong},
  {and} \bibinfo{person}{Jinwook Seo}.} \bibinfo{year}{2021}\natexlab{}.
\newblock \showarticletitle{Papers101: {Supporting} the {Discovery} {Process}
  in the {Literature} {Review} {Workflow} for {Novice} {Researchers}}. In
  \bibinfo{booktitle}{\emph{2021 {IEEE} 14th {Pacific} {Visualization}
  {Symposium} ({PacificVis})}}. \bibinfo{pages}{176--180}.
\newblock
\urldef\tempurl%
\url{https://doi.org/10.1109/PacificVis52677.2021.00037}
\showDOI{\tempurl}
\newblock
\shownote{ISSN: 2165-8773}.


\bibitem[Chou et~al\mbox{.}(2022)]%
        {chou_counterfactuals_2022}
\bibfield{author}{\bibinfo{person}{Yu-Liang Chou}, \bibinfo{person}{Catarina
  Moreira}, \bibinfo{person}{Peter Bruza}, \bibinfo{person}{Chun Ouyang}, {and}
  \bibinfo{person}{Joaquim Jorge}.} \bibinfo{year}{2022}\natexlab{}.
\newblock \showarticletitle{Counterfactuals and causability in explainable
  artificial intelligence: {Theory}, algorithms, and applications}.
\newblock \bibinfo{journal}{\emph{Information Fusion}}  \bibinfo{volume}{81}
  (\bibinfo{date}{May} \bibinfo{year}{2022}), \bibinfo{pages}{59--83}.
\newblock
\showISSN{1566-2535}
\urldef\tempurl%
\url{https://doi.org/10.1016/j.inffus.2021.11.003}
\showDOI{\tempurl}


\bibitem[Cohen(2011)]%
        {cohen2011introduction}
\bibfield{author}{\bibinfo{person}{Morris~F Cohen}.}
  \bibinfo{year}{2011}\natexlab{}.
\newblock \bibinfo{booktitle}{\emph{An introduction to logic and scientific
  method}}.
\newblock \bibinfo{publisher}{Read Books Ltd}.
\newblock


\bibitem[Cole et~al\mbox{.}(2023)]%
        {cole_integrative_2023}
\bibfield{author}{\bibinfo{person}{Nicki~Lisa Cole}, \bibinfo{person}{Sven
  Ulpts}, \bibinfo{person}{Tony Ross-Hellauer}, \bibinfo{person}{Agata
  Bochynska}, {and} \bibinfo{person}{Thomas Klebel}.}
  \bibinfo{year}{2023}\natexlab{}.
\newblock \showarticletitle{Integrative review of conceptions and facilitators
  of and barriers to reproducibility of qualitative research}.
\newblock  (\bibinfo{date}{July} \bibinfo{year}{2023}).
\newblock
\urldef\tempurl%
\url{https://doi.org/10.17605/OSF.IO/Q4XWK}
\showDOI{\tempurl}
\newblock
\shownote{Publisher: OSF}.


\bibitem[Collaboration(2015a)]%
        {collaboration_estimating_2015}
\bibfield{author}{\bibinfo{person}{Open~Science Collaboration}.}
  \bibinfo{year}{2015}\natexlab{a}.
\newblock \showarticletitle{Estimating the reproducibility of psychological
  science}.
\newblock \bibinfo{journal}{\emph{Science}} \bibinfo{volume}{349},
  \bibinfo{number}{6251} (\bibinfo{year}{2015}), \bibinfo{pages}{aac4716}.
\newblock
\newblock
\shownote{ISBN: 0036-8075 Publisher: American Association for the Advancement
  of Science}.


\bibitem[Collaboration(2015b)]%
        {open2015estimating}
\bibfield{author}{\bibinfo{person}{Open~Science Collaboration}.}
  \bibinfo{year}{2015}\natexlab{b}.
\newblock \showarticletitle{Estimating the reproducibility of psychological
  science}.
\newblock \bibinfo{journal}{\emph{Science}} \bibinfo{volume}{349},
  \bibinfo{number}{6251} (\bibinfo{year}{2015}), \bibinfo{pages}{aac4716}.
\newblock


\bibitem[Collberg et~al\mbox{.}(2014)]%
        {collberg2014measuring}
\bibfield{author}{\bibinfo{person}{Christian Collberg}, \bibinfo{person}{Todd
  Proebsting}, \bibinfo{person}{Gina Moraila}, \bibinfo{person}{Akash
  Shankaran}, \bibinfo{person}{Zuoming Shi}, {and} \bibinfo{person}{Alex~M
  Warren}.} \bibinfo{year}{2014}\natexlab{}.
\newblock \showarticletitle{Measuring reproducibility in computer systems
  research}.
\newblock \bibinfo{journal}{\emph{Department of Computer Science, University of
  Arizona, Tech. Rep}}  \bibinfo{volume}{37} (\bibinfo{year}{2014}).
\newblock


\bibitem[Colquhoun(2017)]%
        {colquhoun2017reproducibility}
\bibfield{author}{\bibinfo{person}{David Colquhoun}.}
  \bibinfo{year}{2017}\natexlab{}.
\newblock \showarticletitle{The reproducibility of research and the
  misinterpretation of p-values}.
\newblock \bibinfo{journal}{\emph{Royal society open science}}
  \bibinfo{volume}{4}, \bibinfo{number}{12} (\bibinfo{year}{2017}),
  \bibinfo{pages}{171085}.
\newblock


\bibitem[Cooper et~al\mbox{.}(2018)]%
        {cooper_defining_2018}
\bibfield{author}{\bibinfo{person}{Chris Cooper}, \bibinfo{person}{Andrew
  Booth}, \bibinfo{person}{Jo Varley-Campbell}, \bibinfo{person}{Nicky
  Britten}, {and} \bibinfo{person}{Ruth Garside}.}
  \bibinfo{year}{2018}\natexlab{}.
\newblock \showarticletitle{Defining the process to literature searching in
  systematic reviews: a literature review of guidance and supporting studies}.
\newblock \bibinfo{journal}{\emph{BMC Medical Research Methodology}}
  \bibinfo{volume}{18}, \bibinfo{number}{1} (\bibinfo{date}{Aug.}
  \bibinfo{year}{2018}), \bibinfo{pages}{85}.
\newblock
\showISSN{1471-2288}
\urldef\tempurl%
\url{https://doi.org/10.1186/s12874-018-0545-3}
\showDOI{\tempurl}


\bibitem[Cova et~al\mbox{.}(2021)]%
        {cova2021estimating}
\bibfield{author}{\bibinfo{person}{Florian Cova}, \bibinfo{person}{Brent
  Strickland}, \bibinfo{person}{Angela Abatista}, \bibinfo{person}{Aur{\'e}lien
  Allard}, \bibinfo{person}{James Andow}, \bibinfo{person}{Mario Attie},
  \bibinfo{person}{James Beebe}, \bibinfo{person}{Renatas Berni{\=u}nas},
  \bibinfo{person}{Jordane Boudesseul}, \bibinfo{person}{Matteo Colombo},
  {et~al\mbox{.}}} \bibinfo{year}{2021}\natexlab{}.
\newblock \showarticletitle{Estimating the reproducibility of experimental
  philosophy}.
\newblock \bibinfo{journal}{\emph{Review of Philosophy and Psychology}}
  \bibinfo{volume}{12} (\bibinfo{year}{2021}), \bibinfo{pages}{9--44}.
\newblock


\bibitem[Creswell and Miller(2000)]%
        {creswell2000determining}
\bibfield{author}{\bibinfo{person}{John~W Creswell} {and}
  \bibinfo{person}{Dana~L Miller}.} \bibinfo{year}{2000}\natexlab{}.
\newblock \showarticletitle{Determining validity in qualitative inquiry}.
\newblock \bibinfo{journal}{\emph{Theory into practice}} \bibinfo{volume}{39},
  \bibinfo{number}{3} (\bibinfo{year}{2000}), \bibinfo{pages}{124--130}.
\newblock


\bibitem[Dashnow et~al\mbox{.}(2014)]%
        {dashnow_ten_2014}
\bibfield{author}{\bibinfo{person}{Harriet Dashnow}, \bibinfo{person}{Andrew
  Lonsdale}, {and} \bibinfo{person}{Philip~E. Bourne}.}
  \bibinfo{year}{2014}\natexlab{}.
\newblock \showarticletitle{Ten {Simple} {Rules} for {Writing} a {PLOS} {Ten}
  {Simple} {Rules} {Article}}.
\newblock \bibinfo{journal}{\emph{PLOS Computational Biology}}
  \bibinfo{volume}{10}, \bibinfo{number}{10} (\bibinfo{date}{Oct.}
  \bibinfo{year}{2014}), \bibinfo{pages}{e1003858}.
\newblock
\showISSN{1553-7358}
\urldef\tempurl%
\url{https://doi.org/10.1371/journal.pcbi.1003858}
\showDOI{\tempurl}
\newblock
\shownote{Publisher: Public Library of Science}.


\bibitem[De~Chaumont et~al\mbox{.}(2012)]%
        {de_chaumont_icy_2012}
\bibfield{author}{\bibinfo{person}{Fabrice De~Chaumont},
  \bibinfo{person}{Stéphane Dallongeville}, \bibinfo{person}{Nicolas
  Chenouard}, \bibinfo{person}{Nicolas Hervé}, \bibinfo{person}{Sorin Pop},
  \bibinfo{person}{Thomas Provoost}, \bibinfo{person}{Vannary Meas-Yedid},
  \bibinfo{person}{Praveen Pankajakshan}, \bibinfo{person}{Timothée Lecomte},
  {and} \bibinfo{person}{Yoann Le~Montagner}.} \bibinfo{year}{2012}\natexlab{}.
\newblock \showarticletitle{Icy: an open bioimage informatics platform for
  extended reproducible research}.
\newblock \bibinfo{journal}{\emph{Nature methods}} \bibinfo{volume}{9},
  \bibinfo{number}{7} (\bibinfo{year}{2012}), \bibinfo{pages}{690--696}.
\newblock
\newblock
\shownote{ISBN: 1548-7091 Publisher: Nature Publishing Group US New York}.


\bibitem[Demir et~al\mbox{.}(2022)]%
        {demir_reproducibility_2022}
\bibfield{author}{\bibinfo{person}{Nurullah Demir}, \bibinfo{person}{Matteo
  Große-Kampmann}, \bibinfo{person}{Tobias Urban}, \bibinfo{person}{Christian
  Wressnegger}, \bibinfo{person}{Thorsten Holz}, {and} \bibinfo{person}{Norbert
  Pohlmann}.} \bibinfo{year}{2022}\natexlab{}.
\newblock \showarticletitle{Reproducibility and {Replicability} of {Web}
  {Measurement} {Studies}}. In \bibinfo{booktitle}{\emph{Proceedings of the
  {ACM} {Web} {Conference} 2022}}. \bibinfo{publisher}{ACM},
  \bibinfo{address}{Virtual Event, Lyon France}, \bibinfo{pages}{533--544}.
\newblock
\showISBNx{978-1-4503-9096-5}
\urldef\tempurl%
\url{https://doi.org/10.1145/3485447.3512214}
\showDOI{\tempurl}


\bibitem[Devezer et~al\mbox{.}(2019)]%
        {devezer2019scientific}
\bibfield{author}{\bibinfo{person}{Berna Devezer}, \bibinfo{person}{Luis~G
  Nardin}, \bibinfo{person}{Bert Baumgaertner}, {and}
  \bibinfo{person}{Erkan~Ozge Buzbas}.} \bibinfo{year}{2019}\natexlab{}.
\newblock \showarticletitle{Scientific discovery in a model-centric framework:
  Reproducibility, innovation, and epistemic diversity}.
\newblock \bibinfo{journal}{\emph{PloS one}} \bibinfo{volume}{14},
  \bibinfo{number}{5} (\bibinfo{year}{2019}), \bibinfo{pages}{e0216125}.
\newblock


\bibitem[Ding et~al\mbox{.}(2022)]%
        {ding2022uploaders}
\bibfield{author}{\bibinfo{person}{Xianghua Ding}, \bibinfo{person}{Yubo Kou},
  \bibinfo{person}{Yiwen Xu}, {and} \bibinfo{person}{Peng Zhang}.}
  \bibinfo{year}{2022}\natexlab{}.
\newblock \showarticletitle{“As Uploaders, We Have the Responsibility”:
  Individualized Professionalization of Bilibili Uploaders}. In
  \bibinfo{booktitle}{\emph{Proceedings of the 2022 CHI Conference on Human
  Factors in Computing Systems}}. \bibinfo{pages}{1--14}.
\newblock


\bibitem[Dunne et~al\mbox{.}(2012)]%
        {dunne_rapid_2012}
\bibfield{author}{\bibinfo{person}{Cody Dunne}, \bibinfo{person}{Ben
  Shneiderman}, \bibinfo{person}{Robert Gove}, \bibinfo{person}{Judith
  Klavans}, {and} \bibinfo{person}{Bonnie Dorr}.}
  \bibinfo{year}{2012}\natexlab{}.
\newblock \showarticletitle{Rapid understanding of scientific paper
  collections: {Integrating} statistics, text analytics, and visualization}.
\newblock \bibinfo{journal}{\emph{Journal of the American Society for
  Information Science and Technology}} \bibinfo{volume}{63},
  \bibinfo{number}{12} (\bibinfo{year}{2012}), \bibinfo{pages}{2351--2369}.
\newblock
\showISSN{1532-2890}
\urldef\tempurl%
\url{https://doi.org/10.1002/asi.22652}
\showDOI{\tempurl}
\newblock
\shownote{\_eprint: https://onlinelibrary.wiley.com/doi/pdf/10.1002/asi.22652}.


\bibitem[Durbin(2009)]%
        {durbin_how_2009}
\bibfield{author}{\bibinfo{person}{Charles~G. Durbin}.}
  \bibinfo{year}{2009}\natexlab{}.
\newblock \showarticletitle{How to read a scientific research paper}.
\newblock \bibinfo{journal}{\emph{Respiratory care}} \bibinfo{volume}{54},
  \bibinfo{number}{10} (\bibinfo{year}{2009}), \bibinfo{pages}{1366--1371}.
\newblock
\newblock
\shownote{ISBN: 0020-1324 Publisher: Respiratory Care}.


\bibitem[Ehsan et~al\mbox{.}(2021)]%
        {ehsan2021expanding}
\bibfield{author}{\bibinfo{person}{Upol Ehsan}, \bibinfo{person}{Q~Vera Liao},
  \bibinfo{person}{Michael Muller}, \bibinfo{person}{Mark~O Riedl}, {and}
  \bibinfo{person}{Justin~D Weisz}.} \bibinfo{year}{2021}\natexlab{}.
\newblock \showarticletitle{Expanding explainability: Towards social
  transparency in ai systems}. In \bibinfo{booktitle}{\emph{Proceedings of the
  2021 CHI Conference on Human Factors in Computing Systems}}.
  \bibinfo{pages}{1--19}.
\newblock


\bibitem[Eriksson(2023)]%
        {eriksson2023art}
\bibfield{author}{\bibinfo{person}{David Eriksson}.}
  \bibinfo{year}{2023}\natexlab{}.
\newblock \showarticletitle{The art and science of scholarly writing: framing
  symmetry of specificity beyond IMRAD}.
\newblock \bibinfo{journal}{\emph{European Business Review}}
  \bibinfo{number}{ahead-of-print} (\bibinfo{year}{2023}).
\newblock


\bibitem[Errington et~al\mbox{.}(2014)]%
        {errington2014open}
\bibfield{author}{\bibinfo{person}{Timothy~M Errington},
  \bibinfo{person}{Elizabeth Iorns}, \bibinfo{person}{William Gunn},
  \bibinfo{person}{Fraser~Elisabeth Tan}, \bibinfo{person}{Joelle Lomax}, {and}
  \bibinfo{person}{Brian~A Nosek}.} \bibinfo{year}{2014}\natexlab{}.
\newblock \showarticletitle{An open investigation of the reproducibility of
  cancer biology research}.
\newblock \bibinfo{journal}{\emph{Elife}}  \bibinfo{volume}{3}
  (\bibinfo{year}{2014}), \bibinfo{pages}{e04333}.
\newblock


\bibitem[Fortino et~al\mbox{.}(2020)]%
        {fortino2020using}
\bibfield{author}{\bibinfo{person}{Andres Fortino}, \bibinfo{person}{Qitong
  Zhong}, \bibinfo{person}{Luke Yeh}, {and} \bibinfo{person}{Sijia Fang}.}
  \bibinfo{year}{2020}\natexlab{}.
\newblock \showarticletitle{Using Text Data Mining to Enhance the Literature
  Search Process for Novice STEM Researchers}. In
  \bibinfo{booktitle}{\emph{2020 IEEE Integrated STEM Education Conference
  (ISEC)}}. IEEE, \bibinfo{pages}{1--6}.
\newblock


\bibitem[Foster and Deardorff(2017)]%
        {foster_open_2017}
\bibfield{author}{\bibinfo{person}{Erin~D. Foster} {and} \bibinfo{person}{Ariel
  Deardorff}.} \bibinfo{year}{2017}\natexlab{}.
\newblock \showarticletitle{Open science framework ({OSF})}.
\newblock \bibinfo{journal}{\emph{Journal of the Medical Library Association:
  JMLA}} \bibinfo{volume}{105}, \bibinfo{number}{2} (\bibinfo{year}{2017}),
  \bibinfo{pages}{203}.
\newblock
\newblock
\shownote{Publisher: Medical Library Association}.


\bibitem[Freese et~al\mbox{.}(2022)]%
        {freese2022advances}
\bibfield{author}{\bibinfo{person}{Jeremy Freese}, \bibinfo{person}{Tamkinat
  Rauf}, {and} \bibinfo{person}{Jan~Gerrit Voelkel}.}
  \bibinfo{year}{2022}\natexlab{}.
\newblock \showarticletitle{Advances in transparency and reproducibility in the
  social sciences}.
\newblock \bibinfo{journal}{\emph{Social Science Research}}
  \bibinfo{volume}{107} (\bibinfo{year}{2022}), \bibinfo{pages}{102770}.
\newblock


\bibitem[Furlong and Lester(2023)]%
        {furlong2023toward}
\bibfield{author}{\bibinfo{person}{Darcy~E Furlong} {and}
  \bibinfo{person}{Jessica~Nina Lester}.} \bibinfo{year}{2023}\natexlab{}.
\newblock \showarticletitle{Toward a Practice of Qualitative Methodological
  Literature Reviewing}.
\newblock \bibinfo{journal}{\emph{Qualitative Inquiry}} \bibinfo{volume}{29},
  \bibinfo{number}{6} (\bibinfo{year}{2023}), \bibinfo{pages}{669--677}.
\newblock


\bibitem[Ganguly et~al\mbox{.}(2023)]%
        {ganguly2023review}
\bibfield{author}{\bibinfo{person}{Niloy Ganguly}, \bibinfo{person}{Dren
  Fazlija}, \bibinfo{person}{Maryam Badar}, \bibinfo{person}{Marco Fisichella},
  \bibinfo{person}{Sandipan Sikdar}, \bibinfo{person}{Johanna Schrader},
  \bibinfo{person}{Jonas Wallat}, \bibinfo{person}{Koustav Rudra},
  \bibinfo{person}{Manolis Koubarakis}, \bibinfo{person}{Gourab~K Patro},
  {et~al\mbox{.}}} \bibinfo{year}{2023}\natexlab{}.
\newblock \showarticletitle{A review of the role of causality in developing
  trustworthy ai systems}.
\newblock \bibinfo{journal}{\emph{arXiv preprint arXiv:2302.06975}}
  (\bibinfo{year}{2023}).
\newblock


\bibitem[Garfield(1977)]%
        {garfield_proposal_1977}
\bibfield{author}{\bibinfo{person}{Eugene Garfield}.}
  \bibinfo{year}{1977}\natexlab{}.
\newblock \showarticletitle{Proposal for a new profession-scientific reviewer}.
\newblock \bibinfo{journal}{\emph{Current contents}} \bibinfo{number}{14}
  (\bibinfo{year}{1977}), \bibinfo{pages}{5--8}.
\newblock
\newblock
\shownote{Publisher: INST SCI INFORM INC 3501 MARKET ST, PHILADELPHIA, PA
  19104}.


\bibitem[Greenhalgh(1997)]%
        {greenhalgh_how_1997}
\bibfield{author}{\bibinfo{person}{Trisha Greenhalgh}.}
  \bibinfo{year}{1997}\natexlab{}.
\newblock \showarticletitle{How to read a paper: {Assessing} the methodological
  quality of published papers}.
\newblock \bibinfo{journal}{\emph{Bmj}} \bibinfo{volume}{315},
  \bibinfo{number}{7103} (\bibinfo{year}{1997}), \bibinfo{pages}{305--308}.
\newblock
\newblock
\shownote{ISBN: 0959-8138 Publisher: British Medical Journal Publishing Group}.


\bibitem[Gundersen and Kjensmo(2018)]%
        {gundersen_state_2018}
\bibfield{author}{\bibinfo{person}{Odd~Erik Gundersen} {and}
  \bibinfo{person}{Sigbjørn Kjensmo}.} \bibinfo{year}{2018}\natexlab{}.
\newblock \showarticletitle{State of the art: {Reproducibility} in artificial
  intelligence}. In \bibinfo{booktitle}{\emph{Proceedings of the {AAAI}
  {Conference} on {Artificial} {Intelligence}}}, Vol.~\bibinfo{volume}{32}.
\newblock
\showISBNx{2374-3468}
\newblock
\shownote{Issue: 1}.


\bibitem[Gunning et~al\mbox{.}(2019)]%
        {gunning_xaiexplainable_2019}
\bibfield{author}{\bibinfo{person}{David Gunning}, \bibinfo{person}{Mark
  Stefik}, \bibinfo{person}{Jaesik Choi}, \bibinfo{person}{Timothy Miller},
  \bibinfo{person}{Simone Stumpf}, {and} \bibinfo{person}{Guang-Zhong Yang}.}
  \bibinfo{year}{2019}\natexlab{}.
\newblock \showarticletitle{{XAI}—{Explainable} artificial intelligence}.
\newblock \bibinfo{journal}{\emph{Science Robotics}} \bibinfo{volume}{4},
  \bibinfo{number}{37} (\bibinfo{date}{Dec.} \bibinfo{year}{2019}),
  \bibinfo{pages}{eaay7120}.
\newblock
\showISSN{2470-9476}
\urldef\tempurl%
\url{https://doi.org/10.1126/scirobotics.aay7120}
\showDOI{\tempurl}


\bibitem[Harari et~al\mbox{.}(2020)]%
        {harari_literature_2020}
\bibfield{author}{\bibinfo{person}{Michael~B. Harari},
  \bibinfo{person}{Heather~R. Parola}, \bibinfo{person}{Christopher~J.
  Hartwell}, {and} \bibinfo{person}{Amy Riegelman}.}
  \bibinfo{year}{2020}\natexlab{}.
\newblock \showarticletitle{Literature searches in systematic reviews and
  meta-analyses: {A} review, evaluation, and recommendations}.
\newblock \bibinfo{journal}{\emph{Journal of Vocational Behavior}}
  \bibinfo{volume}{118} (\bibinfo{date}{April} \bibinfo{year}{2020}),
  \bibinfo{pages}{103377}.
\newblock
\showISSN{0001-8791}
\urldef\tempurl%
\url{https://doi.org/10.1016/j.jvb.2020.103377}
\showDOI{\tempurl}


\bibitem[Head et~al\mbox{.}(2021)]%
        {head_augmenting_2021}
\bibfield{author}{\bibinfo{person}{Andrew Head}, \bibinfo{person}{Kyle Lo},
  \bibinfo{person}{Dongyeop Kang}, \bibinfo{person}{Raymond Fok},
  \bibinfo{person}{Sam Skjonsberg}, \bibinfo{person}{Daniel~S. Weld}, {and}
  \bibinfo{person}{Marti~A. Hearst}.} \bibinfo{year}{2021}\natexlab{}.
\newblock \showarticletitle{Augmenting {Scientific} {Papers} with
  {Just}-in-{Time}, {Position}-{Sensitive} {Definitions} of {Terms} and
  {Symbols}}. In \bibinfo{booktitle}{\emph{Proceedings of the 2021 {CHI}
  {Conference} on {Human} {Factors} in {Computing} {Systems}}}
  \emph{(\bibinfo{series}{{CHI} '21})}. \bibinfo{publisher}{Association for
  Computing Machinery}, \bibinfo{address}{New York, NY, USA},
  \bibinfo{pages}{1--18}.
\newblock
\showISBNx{978-1-4503-8096-6}
\urldef\tempurl%
\url{https://doi.org/10.1145/3411764.3445648}
\showDOI{\tempurl}


\bibitem[Head et~al\mbox{.}(2015)]%
        {head_extent_2015}
\bibfield{author}{\bibinfo{person}{Megan~L. Head}, \bibinfo{person}{Luke
  Holman}, \bibinfo{person}{Rob Lanfear}, \bibinfo{person}{Andrew~T. Kahn},
  {and} \bibinfo{person}{Michael~D. Jennions}.}
  \bibinfo{year}{2015}\natexlab{}.
\newblock \showarticletitle{The extent and consequences of p-hacking in
  science}.
\newblock \bibinfo{journal}{\emph{PLoS biology}} \bibinfo{volume}{13},
  \bibinfo{number}{3} (\bibinfo{year}{2015}), \bibinfo{pages}{e1002106}.
\newblock
\newblock
\shownote{ISBN: 1545-7885 Publisher: Public Library of Science}.


\bibitem[Hill-Yardin et~al\mbox{.}(2023)]%
        {hill2023chat}
\bibfield{author}{\bibinfo{person}{Elisa~L Hill-Yardin},
  \bibinfo{person}{Mark~R Hutchinson}, \bibinfo{person}{Robin Laycock}, {and}
  \bibinfo{person}{Sarah~J Spencer}.} \bibinfo{year}{2023}\natexlab{}.
\newblock \showarticletitle{A Chat (GPT) about the future of scientific
  publishing}.
\newblock \bibinfo{journal}{\emph{Brain Behav Immun}}  \bibinfo{volume}{110}
  (\bibinfo{year}{2023}), \bibinfo{pages}{152--154}.
\newblock


\bibitem[Hinderks et~al\mbox{.}(2020)]%
        {hinderks2020slr}
\bibfield{author}{\bibinfo{person}{Andreas Hinderks},
  \bibinfo{person}{Francisco Jos{\'e}~Dom{\'\i}nguez Mayo},
  \bibinfo{person}{J{\"o}rg Thomaschewski}, {and}
  \bibinfo{person}{Mar{\'\i}a~Jos{\'e} Escalona}.}
  \bibinfo{year}{2020}\natexlab{}.
\newblock \showarticletitle{An SLR-tool: Search process in practice: A tool to
  conduct and manage systematic literature review (SLR)}. In
  \bibinfo{booktitle}{\emph{Proceedings of the ACM/IEEE 42nd International
  Conference on Software Engineering: Companion Proceedings}}.
  \bibinfo{pages}{81--84}.
\newblock


\bibitem[Huang et~al\mbox{.}(2020)]%
        {huang2020you}
\bibfield{author}{\bibinfo{person}{Xiaoyun Huang}, \bibinfo{person}{Jessica
  Vitak}, {and} \bibinfo{person}{Yla Tausczik}.}
  \bibinfo{year}{2020}\natexlab{}.
\newblock \showarticletitle{" You Don't Have To Know My Past": How WeChat
  Moments Users Manage Their Evolving Self-Presentation}. In
  \bibinfo{booktitle}{\emph{Proceedings of the 2020 CHI Conference on Human
  Factors in Computing systems}}. \bibinfo{pages}{1--13}.
\newblock


\bibitem[Héroux et~al\mbox{.}(2022)]%
        {heroux_quality_2022}
\bibfield{author}{\bibinfo{person}{Martin~E. Héroux},
  \bibinfo{person}{Annie~A. Butler}, \bibinfo{person}{Aidan~G. Cashin},
  \bibinfo{person}{Euan~J. McCaughey}, \bibinfo{person}{Andrew~J. Affleck},
  \bibinfo{person}{Michael~A. Green}, \bibinfo{person}{Andrew Cartwright},
  \bibinfo{person}{Matthew Jones}, \bibinfo{person}{Kim~M. Kiely},
  \bibinfo{person}{Kimberley S.~van Schooten}, \bibinfo{person}{Jasmine~C.
  Menant}, \bibinfo{person}{Michael Wewege}, {and} \bibinfo{person}{Simon~C.
  Gandevia}.} \bibinfo{year}{2022}\natexlab{}.
\newblock \showarticletitle{Quality {Output} {Checklist} and {Content}
  {Assessment} ({QuOCCA}): a new tool for assessing research quality and
  reproducibility}.
\newblock \bibinfo{journal}{\emph{BMJ Open}} \bibinfo{volume}{12},
  \bibinfo{number}{9} (\bibinfo{date}{Sept.} \bibinfo{year}{2022}),
  \bibinfo{pages}{e060976}.
\newblock
\showISSN{2044-6055, 2044-6055}
\urldef\tempurl%
\url{https://doi.org/10.1136/bmjopen-2022-060976}
\showDOI{\tempurl}
\newblock
\shownote{Publisher: British Medical Journal Publishing Group Section:
  Communication}.


\bibitem[Ince et~al\mbox{.}(2018)]%
        {ince2018study}
\bibfield{author}{\bibinfo{person}{Sharon~Favaro Ince},
  \bibinfo{person}{Christopher Hoadley}, {and} \bibinfo{person}{Paul~A
  Kirschner}.} \bibinfo{year}{2018}\natexlab{}.
\newblock \showarticletitle{A study of search practices in doctoral student
  scholarly workflows}. In \bibinfo{booktitle}{\emph{Proceedings of the 2018
  Conference on Human Information Interaction \& Retrieval}}.
  \bibinfo{pages}{245--248}.
\newblock


\bibitem[Keshav(2007)]%
        {keshav_how_2007}
\bibfield{author}{\bibinfo{person}{Srinivasan Keshav}.}
  \bibinfo{year}{2007}\natexlab{}.
\newblock \showarticletitle{How to read a paper}.
\newblock \bibinfo{journal}{\emph{ACM SIGCOMM Computer Communication Review}}
  \bibinfo{volume}{37}, \bibinfo{number}{3} (\bibinfo{year}{2007}),
  \bibinfo{pages}{83--84}.
\newblock
\newblock
\shownote{ISBN: 0146-4833 Publisher: ACM New York, NY, USA}.


\bibitem[Kidwell et~al\mbox{.}(2016)]%
        {kidwell_badges_2016}
\bibfield{author}{\bibinfo{person}{Mallory~C. Kidwell},
  \bibinfo{person}{Ljiljana~B. Lazarević}, \bibinfo{person}{Erica Baranski},
  \bibinfo{person}{Tom~E. Hardwicke}, \bibinfo{person}{Sarah Piechowski},
  \bibinfo{person}{Lina-Sophia Falkenberg}, \bibinfo{person}{Curtis Kennett},
  \bibinfo{person}{Agnieszka Slowik}, \bibinfo{person}{Carina Sonnleitner},
  {and} \bibinfo{person}{Chelsey Hess-Holden}.}
  \bibinfo{year}{2016}\natexlab{}.
\newblock \showarticletitle{Badges to acknowledge open practices: {A} simple,
  low-cost, effective method for increasing transparency}.
\newblock \bibinfo{journal}{\emph{PLoS biology}} \bibinfo{volume}{14},
  \bibinfo{number}{5} (\bibinfo{year}{2016}), \bibinfo{pages}{e1002456}.
\newblock
\newblock
\shownote{ISBN: 1544-9173 Publisher: Public Library of Science San Francisco,
  CA USA}.


\bibitem[Kim et~al\mbox{.}(2023)]%
        {kim2023help}
\bibfield{author}{\bibinfo{person}{Sunnie~SY Kim},
  \bibinfo{person}{Elizabeth~Anne Watkins}, \bibinfo{person}{Olga Russakovsky},
  \bibinfo{person}{Ruth Fong}, {and} \bibinfo{person}{Andr{\'e}s
  Monroy-Hern{\'a}ndez}.} \bibinfo{year}{2023}\natexlab{}.
\newblock \showarticletitle{" Help Me Help the AI": Understanding How
  Explainability Can Support Human-AI Interaction}. In
  \bibinfo{booktitle}{\emph{Proceedings of the 2023 CHI Conference on Human
  Factors in Computing Systems}}. \bibinfo{pages}{1--17}.
\newblock


\bibitem[King et~al\mbox{.}(2020)]%
        {king_search_2020}
\bibfield{author}{\bibinfo{person}{Seth~A. King}, \bibinfo{person}{Douglas
  Kostewicz}, \bibinfo{person}{Olivia Enders}, \bibinfo{person}{Taneal Burch},
  \bibinfo{person}{Argnue Chitiyo}, \bibinfo{person}{Johanna Taylor},
  \bibinfo{person}{Sarah DeMaria}, {and} \bibinfo{person}{Milsha Reid}.}
  \bibinfo{year}{2020}\natexlab{}.
\newblock \showarticletitle{Search and {Selection} {Procedures} of {Literature}
  {Reviews} in {Behavior} {Analysis}}.
\newblock \bibinfo{journal}{\emph{Perspectives on Behavior Science}}
  \bibinfo{volume}{43}, \bibinfo{number}{4} (\bibinfo{date}{Dec.}
  \bibinfo{year}{2020}), \bibinfo{pages}{725--760}.
\newblock
\showISSN{2520-8977}
\urldef\tempurl%
\url{https://doi.org/10.1007/s40614-020-00265-9}
\showDOI{\tempurl}


\bibitem[Koneru et~al\mbox{.}(2023)]%
        {koneru2023can}
\bibfield{author}{\bibinfo{person}{Sai Koneru}, \bibinfo{person}{Jian Wu},
  {and} \bibinfo{person}{Sarah Rajtmajer}.} \bibinfo{year}{2023}\natexlab{}.
\newblock \showarticletitle{Can Large Language Models Discern Evidence for
  Scientific Hypotheses? Case Studies in the Social Sciences}.
\newblock \bibinfo{journal}{\emph{arXiv preprint arXiv:2309.06578}}
  (\bibinfo{year}{2023}).
\newblock


\bibitem[Kung(2023)]%
        {kung_elicit_2023}
\bibfield{author}{\bibinfo{person}{Janice~Y. Kung}.}
  \bibinfo{year}{2023}\natexlab{}.
\newblock \showarticletitle{Elicit}.
\newblock \bibinfo{journal}{\emph{The Journal of the Canadian Health Libraries
  Association}} \bibinfo{volume}{44}, \bibinfo{number}{1}
  (\bibinfo{date}{April} \bibinfo{year}{2023}), \bibinfo{pages}{15--18}.
\newblock
\showISSN{1708-6892}
\urldef\tempurl%
\url{https://doi.org/10.29173/jchla29657}
\showDOI{\tempurl}


\bibitem[L.~Haven and Van~Grootel(2019)]%
        {l_haven_preregistering_2019}
\bibfield{author}{\bibinfo{person}{Tamarinde L.~Haven} {and}
  \bibinfo{person}{Dr.~Leonie Van~Grootel}.} \bibinfo{year}{2019}\natexlab{}.
\newblock \showarticletitle{Preregistering qualitative research}.
\newblock \bibinfo{journal}{\emph{Accountability in Research}}
  \bibinfo{volume}{26}, \bibinfo{number}{3} (\bibinfo{date}{April}
  \bibinfo{year}{2019}), \bibinfo{pages}{229--244}.
\newblock
\showISSN{0898-9621}
\urldef\tempurl%
\url{https://doi.org/10.1080/08989621.2019.1580147}
\showDOI{\tempurl}
\newblock
\shownote{Publisher: Taylor \& Francis \_eprint:
  https://doi.org/10.1080/08989621.2019.1580147}.


\bibitem[Lalu et~al\mbox{.}(2017)]%
        {lalu2017stakeholders}
\bibfield{author}{\bibinfo{person}{Manoj~Mathew Lalu}, \bibinfo{person}{Larissa
  Shamseer}, \bibinfo{person}{Kelly~D Cobey}, {and} \bibinfo{person}{David
  Moher}.} \bibinfo{year}{2017}\natexlab{}.
\newblock \showarticletitle{How stakeholders can respond to the rise of
  predatory journals}.
\newblock \bibinfo{journal}{\emph{Nature Human Behaviour}} \bibinfo{volume}{1},
  \bibinfo{number}{12} (\bibinfo{year}{2017}), \bibinfo{pages}{852--855}.
\newblock


\bibitem[Leonelli(2018)]%
        {leonelli_rethinking_2018}
\bibfield{author}{\bibinfo{person}{Sabina Leonelli}.}
  \bibinfo{year}{2018}\natexlab{}.
\newblock \showarticletitle{Rethinking {Reproducibility} as a {Criterion} for
  {Research} {Quality}}.
\newblock In \bibinfo{booktitle}{\emph{Including a {Symposium} on {Mary}
  {Morgan}: {Curiosity}, {Imagination}, and {Surprise}}}.
  \bibinfo{series}{Research in the {History} of {Economic} {Thought} and
  {Methodology}}, Vol.~\bibinfo{volume}{36B}. \bibinfo{publisher}{Emerald
  Publishing Limited}, \bibinfo{pages}{129--146}.
\newblock
\showISBNx{978-1-78756-423-7 978-1-78756-424-4}
\urldef\tempurl%
\url{https://doi.org/10.1108/S0743-41542018000036B009}
\showDOI{\tempurl}


\bibitem[Linardatos et~al\mbox{.}(2021)]%
        {linardatos_explainable_2021}
\bibfield{author}{\bibinfo{person}{Pantelis Linardatos},
  \bibinfo{person}{Vasilis Papastefanopoulos}, {and} \bibinfo{person}{Sotiris
  Kotsiantis}.} \bibinfo{year}{2021}\natexlab{}.
\newblock \showarticletitle{Explainable {AI}: {A} {Review} of {Machine}
  {Learning} {Interpretability} {Methods}}.
\newblock \bibinfo{journal}{\emph{Entropy}} \bibinfo{volume}{23},
  \bibinfo{number}{1} (\bibinfo{date}{Jan.} \bibinfo{year}{2021}),
  \bibinfo{pages}{18}.
\newblock
\showISSN{1099-4300}
\urldef\tempurl%
\url{https://doi.org/10.3390/e23010018}
\showDOI{\tempurl}
\newblock
\shownote{Number: 1 Publisher: Multidisciplinary Digital Publishing Institute}.


\bibitem[Liu et~al\mbox{.}(2023)]%
        {liu2023evaluating}
\bibfield{author}{\bibinfo{person}{Nelson~F Liu}, \bibinfo{person}{Tianyi
  Zhang}, {and} \bibinfo{person}{Percy Liang}.}
  \bibinfo{year}{2023}\natexlab{}.
\newblock \showarticletitle{Evaluating verifiability in generative search
  engines}.
\newblock \bibinfo{journal}{\emph{arXiv preprint arXiv:2304.09848}}
  (\bibinfo{year}{2023}).
\newblock


\bibitem[Liu et~al\mbox{.}(2014)]%
        {liu_literature_2014}
\bibfield{author}{\bibinfo{person}{Shengbo Liu}, \bibinfo{person}{Chaomei
  Chen}, \bibinfo{person}{Kun Ding}, \bibinfo{person}{Bo Wang},
  \bibinfo{person}{Kan Xu}, {and} \bibinfo{person}{Yuan Lin}.}
  \bibinfo{year}{2014}\natexlab{}.
\newblock \showarticletitle{Literature retrieval based on citation context}.
\newblock \bibinfo{journal}{\emph{Scientometrics}} \bibinfo{volume}{101},
  \bibinfo{number}{2} (\bibinfo{date}{Nov.} \bibinfo{year}{2014}),
  \bibinfo{pages}{1293--1307}.
\newblock
\showISSN{1588-2861}
\urldef\tempurl%
\url{https://doi.org/10.1007/s11192-014-1233-7}
\showDOI{\tempurl}


\bibitem[Makel et~al\mbox{.}(2012)]%
        {makel_replications_2012}
\bibfield{author}{\bibinfo{person}{Matthew~C. Makel},
  \bibinfo{person}{Jonathan~A. Plucker}, {and} \bibinfo{person}{Boyd Hegarty}.}
  \bibinfo{year}{2012}\natexlab{}.
\newblock \showarticletitle{Replications in psychology research: {How} often do
  they really occur?}
\newblock \bibinfo{journal}{\emph{Perspectives on Psychological Science}}
  \bibinfo{volume}{7}, \bibinfo{number}{6} (\bibinfo{year}{2012}),
  \bibinfo{pages}{537--542}.
\newblock
\newblock
\shownote{ISBN: 1745-6916 Publisher: Sage Publications Sage CA: Los Angeles,
  CA}.


\bibitem[Maloney and Conrad(2016)]%
        {maloney2016expecting}
\bibfield{author}{\bibinfo{person}{Alan Maloney} {and}
  \bibinfo{person}{Lettie~Y Conrad}.} \bibinfo{year}{2016}\natexlab{}.
\newblock \showarticletitle{Expecting the unexpected: Serendipity, discovery,
  and the scholarly research process}.
\newblock \bibinfo{journal}{\emph{White Paper}} (\bibinfo{year}{2016}).
\newblock


\bibitem[Mao et~al\mbox{.}(2018)]%
        {mao2018does}
\bibfield{author}{\bibinfo{person}{Jiaxin Mao}, \bibinfo{person}{Yiqun Liu},
  \bibinfo{person}{Noriko Kando}, \bibinfo{person}{Min Zhang}, {and}
  \bibinfo{person}{Shaoping Ma}.} \bibinfo{year}{2018}\natexlab{}.
\newblock \showarticletitle{How does domain expertise affect users’ search
  interaction and outcome in exploratory search?}
\newblock \bibinfo{journal}{\emph{ACM Transactions on Information Systems
  (TOIS)}} \bibinfo{volume}{36}, \bibinfo{number}{4} (\bibinfo{year}{2018}),
  \bibinfo{pages}{1--30}.
\newblock


\bibitem[Martín-Martín et~al\mbox{.}(2018)]%
        {martin-martin_evidence_2018}
\bibfield{author}{\bibinfo{person}{Alberto Martín-Martín},
  \bibinfo{person}{Rodrigo Costas}, \bibinfo{person}{Thed Van~Leeuwen}, {and}
  \bibinfo{person}{Emilio~Delgado López-Cózar}.}
  \bibinfo{year}{2018}\natexlab{}.
\newblock \showarticletitle{Evidence of open access of scientific publications
  in {Google} {Scholar}: {A} large-scale analysis}.
\newblock \bibinfo{journal}{\emph{Journal of informetrics}}
  \bibinfo{volume}{12}, \bibinfo{number}{3} (\bibinfo{year}{2018}),
  \bibinfo{pages}{819--841}.
\newblock
\newblock
\shownote{ISBN: 1751-1577 Publisher: Elsevier}.


\bibitem[McKiernan et~al\mbox{.}(2016)]%
        {mckiernan_how_2016}
\bibfield{author}{\bibinfo{person}{Erin~C. McKiernan},
  \bibinfo{person}{Philip~E. Bourne}, \bibinfo{person}{C.~Titus Brown},
  \bibinfo{person}{Stuart Buck}, \bibinfo{person}{Amye Kenall},
  \bibinfo{person}{Jennifer Lin}, \bibinfo{person}{Damon McDougall},
  \bibinfo{person}{Brian~A. Nosek}, \bibinfo{person}{Karthik Ram}, {and}
  \bibinfo{person}{Courtney~K. Soderberg}.} \bibinfo{year}{2016}\natexlab{}.
\newblock \showarticletitle{How open science helps researchers succeed}.
\newblock \bibinfo{journal}{\emph{elife}}  \bibinfo{volume}{5}
  (\bibinfo{year}{2016}), \bibinfo{pages}{e16800}.
\newblock
\newblock
\shownote{ISBN: 2050-084X Publisher: eLife Sciences Publications, Ltd}.


\bibitem[McShane et~al\mbox{.}(2019)]%
        {mcshane2019large}
\bibfield{author}{\bibinfo{person}{Blakeley~B McShane},
  \bibinfo{person}{Jennifer~L Tackett}, \bibinfo{person}{Ulf B{\"o}ckenholt},
  {and} \bibinfo{person}{Andrew Gelman}.} \bibinfo{year}{2019}\natexlab{}.
\newblock \showarticletitle{Large-scale replication projects in contemporary
  psychological research}.
\newblock \bibinfo{journal}{\emph{The American Statistician}}
  \bibinfo{volume}{73}, \bibinfo{number}{sup1} (\bibinfo{year}{2019}),
  \bibinfo{pages}{99--105}.
\newblock


\bibitem[Mengist et~al\mbox{.}(2020)]%
        {mengist2020method}
\bibfield{author}{\bibinfo{person}{Wondimagegn Mengist},
  \bibinfo{person}{Teshome Soromessa}, {and} \bibinfo{person}{Gudina Legese}.}
  \bibinfo{year}{2020}\natexlab{}.
\newblock \showarticletitle{Method for conducting systematic literature review
  and meta-analysis for environmental science research}.
\newblock \bibinfo{journal}{\emph{MethodsX}}  \bibinfo{volume}{7}
  (\bibinfo{year}{2020}), \bibinfo{pages}{100777}.
\newblock


\bibitem[Miłkowski et~al\mbox{.}(2018)]%
        {milkowski_replicability_2018}
\bibfield{author}{\bibinfo{person}{Marcin Miłkowski},
  \bibinfo{person}{Witold~M. Hensel}, {and} \bibinfo{person}{Mateusz Hohol}.}
  \bibinfo{year}{2018}\natexlab{}.
\newblock \showarticletitle{Replicability or reproducibility? {On} the
  replication crisis in computational neuroscience and sharing only relevant
  detail}.
\newblock \bibinfo{journal}{\emph{Journal of Computational Neuroscience}}
  \bibinfo{volume}{45}, \bibinfo{number}{3} (\bibinfo{date}{Dec.}
  \bibinfo{year}{2018}), \bibinfo{pages}{163--172}.
\newblock
\showISSN{1573-6873}
\urldef\tempurl%
\url{https://doi.org/10.1007/s10827-018-0702-z}
\showDOI{\tempurl}


\bibitem[Munafò et~al\mbox{.}(2017)]%
        {munafo_manifesto_2017}
\bibfield{author}{\bibinfo{person}{Marcus~R. Munafò},
  \bibinfo{person}{Brian~A. Nosek}, \bibinfo{person}{Dorothy~VM Bishop},
  \bibinfo{person}{Katherine~S. Button}, \bibinfo{person}{Christopher~D.
  Chambers}, \bibinfo{person}{Nathalie Percie~du Sert}, \bibinfo{person}{Uri
  Simonsohn}, \bibinfo{person}{Eric-Jan Wagenmakers},
  \bibinfo{person}{Jennifer~J. Ware}, {and} \bibinfo{person}{John Ioannidis}.}
  \bibinfo{year}{2017}\natexlab{}.
\newblock \showarticletitle{A manifesto for reproducible science}.
\newblock \bibinfo{journal}{\emph{Nature human behaviour}} \bibinfo{volume}{1},
  \bibinfo{number}{1} (\bibinfo{year}{2017}), \bibinfo{pages}{1--9}.
\newblock
\newblock
\shownote{ISBN: 2397-3374 Publisher: Nature Publishing Group}.


\bibitem[Nakshatri et~al\mbox{.}(2021)]%
        {nakshatri2021design}
\bibfield{author}{\bibinfo{person}{Nishanth Nakshatri}, \bibinfo{person}{Arjun
  Menon}, \bibinfo{person}{C~Lee Giles}, \bibinfo{person}{Sarah Rajtmajer},
  {and} \bibinfo{person}{Christopher Griffin}.}
  \bibinfo{year}{2021}\natexlab{}.
\newblock \showarticletitle{Design and Analysis of a Synthetic Prediction
  Market using Dynamic Convex Sets}.
\newblock \bibinfo{journal}{\emph{arXiv preprint arXiv:2101.01787}}
  (\bibinfo{year}{2021}).
\newblock


\bibitem[{National Academies of Sciences, Engineering, and Medicine}
  et~al\mbox{.}(2019)]%
        {national2019reproducibility}
\bibfield{author}{\bibinfo{person}{{National Academies of Sciences,
  Engineering, and Medicine}} {et~al\mbox{.}}} \bibinfo{year}{2019}\natexlab{}.
\newblock \bibinfo{booktitle}{\emph{Reproducibility and replicability in
  science}}.
\newblock \bibinfo{publisher}{National Academies Press}.
\newblock


\bibitem[Nosek et~al\mbox{.}(2021)]%
        {nosek2021replicability}
\bibfield{author}{\bibinfo{person}{Brian~A Nosek}, \bibinfo{person}{Tom~E
  Hardwicke}, \bibinfo{person}{Hannah Moshontz}, \bibinfo{person}{Aur{\'e}lien
  Allard}, \bibinfo{person}{Katherine~S Corker}, \bibinfo{person}{Anna~Dreber
  Almenberg}, \bibinfo{person}{Fiona Fidler}, \bibinfo{person}{Joseph Hilgard},
  \bibinfo{person}{Melissa Kline}, \bibinfo{person}{Mich{\`e}le~B Nuijten},
  {et~al\mbox{.}}} \bibinfo{year}{2021}\natexlab{}.
\newblock \showarticletitle{Replicability, robustness, and reproducibility in
  psychological science}.
\newblock  (\bibinfo{year}{2021}).
\newblock


\bibitem[Page et~al\mbox{.}(2021)]%
        {page2021prisma}
\bibfield{author}{\bibinfo{person}{Matthew~J Page}, \bibinfo{person}{Joanne~E
  McKenzie}, \bibinfo{person}{Patrick~M Bossuyt}, \bibinfo{person}{Isabelle
  Boutron}, \bibinfo{person}{Tammy~C Hoffmann}, \bibinfo{person}{Cynthia~D
  Mulrow}, \bibinfo{person}{Larissa Shamseer}, \bibinfo{person}{Jennifer~M
  Tetzlaff}, \bibinfo{person}{Elie~A Akl}, \bibinfo{person}{Sue~E Brennan},
  {et~al\mbox{.}}} \bibinfo{year}{2021}\natexlab{}.
\newblock \showarticletitle{The PRISMA 2020 statement: an updated guideline for
  reporting systematic reviews}.
\newblock \bibinfo{journal}{\emph{International journal of surgery}}
  \bibinfo{volume}{88} (\bibinfo{year}{2021}), \bibinfo{pages}{105906}.
\newblock


\bibitem[Palani et~al\mbox{.}(2023)]%
        {palani_relatedly_2023}
\bibfield{author}{\bibinfo{person}{Srishti Palani}, \bibinfo{person}{Aakanksha
  Naik}, \bibinfo{person}{Doug Downey}, \bibinfo{person}{Amy~X. Zhang},
  \bibinfo{person}{Jonathan Bragg}, {and} \bibinfo{person}{Joseph~Chee Chang}.}
  \bibinfo{year}{2023}\natexlab{}.
\newblock \showarticletitle{Relatedly: {Scaffolding} {Literature} {Reviews}
  with {Existing} {Related} {Work} {Sections}}. In
  \bibinfo{booktitle}{\emph{Proceedings of the 2023 {CHI} {Conference} on
  {Human} {Factors} in {Computing} {Systems}}}. \bibinfo{publisher}{ACM},
  \bibinfo{address}{Hamburg Germany}, \bibinfo{pages}{1--20}.
\newblock
\showISBNx{978-1-4503-9421-5}
\urldef\tempurl%
\url{https://doi.org/10.1145/3544548.3580841}
\showDOI{\tempurl}


\bibitem[Pawel and Held(2020)]%
        {pawel2020probabilistic}
\bibfield{author}{\bibinfo{person}{Samuel Pawel} {and}
  \bibinfo{person}{Leonhard Held}.} \bibinfo{year}{2020}\natexlab{}.
\newblock \showarticletitle{Probabilistic forecasting of replication studies}.
\newblock \bibinfo{journal}{\emph{PloS one}} \bibinfo{volume}{15},
  \bibinfo{number}{4} (\bibinfo{year}{2020}), \bibinfo{pages}{e0231416}.
\newblock


\bibitem[Pineau et~al\mbox{.}(2021)]%
        {pineau2021improving}
\bibfield{author}{\bibinfo{person}{Joelle Pineau}, \bibinfo{person}{Philippe
  Vincent-Lamarre}, \bibinfo{person}{Koustuv Sinha}, \bibinfo{person}{Vincent
  Larivi{\`e}re}, \bibinfo{person}{Alina Beygelzimer},
  \bibinfo{person}{Florence d’Alch{\'e} Buc}, \bibinfo{person}{Emily Fox},
  {and} \bibinfo{person}{Hugo Larochelle}.} \bibinfo{year}{2021}\natexlab{}.
\newblock \showarticletitle{Improving Reproducibility in Machine Learning
  Research}.
\newblock \bibinfo{journal}{\emph{Journal of Machine Learning Research}}
  \bibinfo{volume}{22} (\bibinfo{year}{2021}).
\newblock


\bibitem[Plesser(2018)]%
        {plesser_reproducibility_2018}
\bibfield{author}{\bibinfo{person}{Hans~E. Plesser}.}
  \bibinfo{year}{2018}\natexlab{}.
\newblock \showarticletitle{Reproducibility vs. {Replicability}: {A} {Brief}
  {History} of a {Confused} {Terminology}}.
\newblock \bibinfo{journal}{\emph{Frontiers in Neuroinformatics}}
  \bibinfo{volume}{11} (\bibinfo{year}{2018}).
\newblock
\showISSN{1662-5196}
\urldef\tempurl%
\url{https://www.frontiersin.org/articles/10.3389/fninf.2017.00076}
\showURL{%
\tempurl}


\bibitem[Ponte and Simon(2011)]%
        {ponte_scholarly_2011}
\bibfield{author}{\bibinfo{person}{Diego Ponte} {and} \bibinfo{person}{Judith
  Simon}.} \bibinfo{year}{2011}\natexlab{}.
\newblock \showarticletitle{Scholarly {Communication} 2.0: {Exploring}
  {Researchers}' {Opinions} on {Web} 2.0 for {Scientific} {Knowledge}
  {Creation}, {Evaluation} and {Dissemination}}.
\newblock \bibinfo{journal}{\emph{Serials Review}} \bibinfo{volume}{37},
  \bibinfo{number}{3} (\bibinfo{date}{Sept.} \bibinfo{year}{2011}),
  \bibinfo{pages}{149--156}.
\newblock
\showISSN{0098-7913}
\urldef\tempurl%
\url{https://doi.org/10.1080/00987913.2011.10765376}
\showDOI{\tempurl}
\newblock
\shownote{Publisher: Routledge \_eprint:
  https://www.tandfonline.com/doi/pdf/10.1080/00987913.2011.10765376}.


\bibitem[Prabhudesai et~al\mbox{.}(2023)]%
        {prabhudesai2023understanding}
\bibfield{author}{\bibinfo{person}{Snehal Prabhudesai}, \bibinfo{person}{Leyao
  Yang}, \bibinfo{person}{Sumit Asthana}, \bibinfo{person}{Xun Huan},
  \bibinfo{person}{Q~Vera Liao}, {and} \bibinfo{person}{Nikola Banovic}.}
  \bibinfo{year}{2023}\natexlab{}.
\newblock \showarticletitle{Understanding Uncertainty: How Lay Decision-makers
  Perceive and Interpret Uncertainty in Human-AI Decision Making}. In
  \bibinfo{booktitle}{\emph{Proceedings of the 28th International Conference on
  Intelligent User Interfaces}}. \bibinfo{pages}{379--396}.
\newblock


\bibitem[Raff(2019)]%
        {raff2019step}
\bibfield{author}{\bibinfo{person}{Edward Raff}.}
  \bibinfo{year}{2019}\natexlab{}.
\newblock \showarticletitle{A step toward quantifying independently
  reproducible machine learning research}.
\newblock \bibinfo{journal}{\emph{Advances in Neural Information Processing
  Systems}}  \bibinfo{volume}{32} (\bibinfo{year}{2019}).
\newblock


\bibitem[Rajtmajer et~al\mbox{.}(2021)]%
        {rajtmajer2021synthetic}
\bibfield{author}{\bibinfo{person}{Sarah Rajtmajer},
  \bibinfo{person}{Christopher Griffin}, \bibinfo{person}{Jian Wu},
  \bibinfo{person}{Robert Fraleigh}, \bibinfo{person}{Laxmaan Balaji},
  \bibinfo{person}{Anna Squicciarini}, \bibinfo{person}{Anthony Kwasnica},
  \bibinfo{person}{David Pennock}, \bibinfo{person}{Michael McLaughlin},
  \bibinfo{person}{Timothy Fritton}, {et~al\mbox{.}}}
  \bibinfo{year}{2021}\natexlab{}.
\newblock \showarticletitle{A Synthetic Prediction Market for Estimating
  Confidence in Published Work}.
\newblock \bibinfo{journal}{\emph{arXiv preprint arXiv:2201.06924}}
  (\bibinfo{year}{2021}).
\newblock


\bibitem[Rajtmajer et~al\mbox{.}(2022)]%
        {rajtmajer2022synthetic}
\bibfield{author}{\bibinfo{person}{Sarah Rajtmajer},
  \bibinfo{person}{Christopher Griffin}, \bibinfo{person}{Jian Wu},
  \bibinfo{person}{Robert Fraleigh}, \bibinfo{person}{Laxmaan Balaji},
  \bibinfo{person}{Anna Squicciarini}, \bibinfo{person}{Anthony Kwasnica},
  \bibinfo{person}{David Pennock}, \bibinfo{person}{Michael McLaughlin},
  \bibinfo{person}{Timothy Fritton}, {et~al\mbox{.}}}
  \bibinfo{year}{2022}\natexlab{}.
\newblock \showarticletitle{A synthetic prediction market for estimating
  confidence in published work}. In \bibinfo{booktitle}{\emph{Proceedings of
  the AAAI Conference on Artificial Intelligence}}, Vol.~\bibinfo{volume}{36}.
  \bibinfo{pages}{13218--13220}.
\newblock


\bibitem[Rethlefsen et~al\mbox{.}(2021)]%
        {rethlefsen2021prisma}
\bibfield{author}{\bibinfo{person}{Melissa~L Rethlefsen},
  \bibinfo{person}{Shona Kirtley}, \bibinfo{person}{Siw Waffenschmidt},
  \bibinfo{person}{Ana~Patricia Ayala}, \bibinfo{person}{David Moher},
  \bibinfo{person}{Matthew~J Page}, {and} \bibinfo{person}{Jonathan~B Koffel}.}
  \bibinfo{year}{2021}\natexlab{}.
\newblock \showarticletitle{PRISMA-S: an extension to the PRISMA statement for
  reporting literature searches in systematic reviews}.
\newblock \bibinfo{journal}{\emph{Systematic reviews}}  \bibinfo{volume}{10}
  (\bibinfo{year}{2021}), \bibinfo{pages}{1--19}.
\newblock


\bibitem[Samek and Müller(2019)]%
        {samek_towards_2019}
\bibfield{author}{\bibinfo{person}{Wojciech Samek} {and}
  \bibinfo{person}{Klaus-Robert Müller}.} \bibinfo{year}{2019}\natexlab{}.
\newblock \showarticletitle{Towards {Explainable} {Artificial} {Intelligence}}.
\newblock In \bibinfo{booktitle}{\emph{Explainable {AI}: {Interpreting},
  {Explaining} and {Visualizing} {Deep} {Learning}}},
  \bibfield{editor}{\bibinfo{person}{Wojciech Samek},
  \bibinfo{person}{Grégoire Montavon}, \bibinfo{person}{Andrea Vedaldi},
  \bibinfo{person}{Lars~Kai Hansen}, {and} \bibinfo{person}{Klaus-Robert
  Müller}} (Eds.). \bibinfo{publisher}{Springer International Publishing},
  \bibinfo{address}{Cham}, \bibinfo{pages}{5--22}.
\newblock
\showISBNx{978-3-030-28954-6}
\urldef\tempurl%
\url{https://doi.org/10.1007/978-3-030-28954-6_1}
\showDOI{\tempurl}


\bibitem[Shiely et~al\mbox{.}(2024)]%
        {shiely2024and}
\bibfield{author}{\bibinfo{person}{Frances Shiely}, \bibinfo{person}{Kerrie
  Gallagher}, {and} \bibinfo{person}{Se{\'a}n~R Millar}.}
  \bibinfo{year}{2024}\natexlab{}.
\newblock \showarticletitle{How, and why, science and health researchers read
  scientific (IMRAD) papers}.
\newblock \bibinfo{journal}{\emph{Plos one}} \bibinfo{volume}{19},
  \bibinfo{number}{1} (\bibinfo{year}{2024}), \bibinfo{pages}{e0297034}.
\newblock


\bibitem[Sollaci and Pereira(2004)]%
        {sollaci_introduction_2004}
\bibfield{author}{\bibinfo{person}{Luciana~B. Sollaci} {and}
  \bibinfo{person}{Mauricio~G. Pereira}.} \bibinfo{year}{2004}\natexlab{}.
\newblock \showarticletitle{The introduction, methods, results, and discussion
  ({IMRAD}) structure: a fifty-year survey}.
\newblock \bibinfo{journal}{\emph{Journal of the medical library association}}
  \bibinfo{volume}{92}, \bibinfo{number}{3} (\bibinfo{year}{2004}),
  \bibinfo{pages}{364}.
\newblock
\newblock
\shownote{Publisher: Medical Library Association}.


\bibitem[Sorokowski et~al\mbox{.}(2017)]%
        {sorokowski2017predatory}
\bibfield{author}{\bibinfo{person}{Piotr Sorokowski}, \bibinfo{person}{Emanuel
  Kulczycki}, \bibinfo{person}{Agnieszka Sorokowska}, {and}
  \bibinfo{person}{Katarzyna Pisanski}.} \bibinfo{year}{2017}\natexlab{}.
\newblock \showarticletitle{Predatory journals recruit fake editor}.
\newblock \bibinfo{journal}{\emph{Nature}} \bibinfo{volume}{543},
  \bibinfo{number}{7646} (\bibinfo{year}{2017}), \bibinfo{pages}{481--483}.
\newblock


\bibitem[So{\'s}nicki and Madeyski(2021)]%
        {sosnicki2021ash}
\bibfield{author}{\bibinfo{person}{Marek So{\'s}nicki} {and}
  \bibinfo{person}{Lech Madeyski}.} \bibinfo{year}{2021}\natexlab{}.
\newblock \showarticletitle{ASH: A New Tool for Automated and Full-Text Search
  in Systematic Literature Reviews}. In \bibinfo{booktitle}{\emph{International
  Conference on Computational Science}}. Springer, \bibinfo{pages}{362--369}.
\newblock


\bibitem[Soufan et~al\mbox{.}(2022)]%
        {soufan2022searching}
\bibfield{author}{\bibinfo{person}{Ayah Soufan}, \bibinfo{person}{Ian Ruthven},
  {and} \bibinfo{person}{Leif Azzopardi}.} \bibinfo{year}{2022}\natexlab{}.
\newblock \showarticletitle{Searching the literature: an analysis of an
  exploratory search task}. In \bibinfo{booktitle}{\emph{Proceedings of the
  2022 Conference on Human Information Interaction and Retrieval}}.
  \bibinfo{pages}{146--157}.
\newblock


\bibitem[Sultanum et~al\mbox{.}(2020)]%
        {sultanum_understanding_2020}
\bibfield{author}{\bibinfo{person}{Nicole Sultanum}, \bibinfo{person}{Christine
  Murad}, {and} \bibinfo{person}{Daniel Wigdor}.}
  \bibinfo{year}{2020}\natexlab{}.
\newblock \showarticletitle{Understanding and {Supporting} {Academic}
  {Literature} {Review} {Workflows} with {LitSense}}. In
  \bibinfo{booktitle}{\emph{Proceedings of the {International} {Conference} on
  {Advanced} {Visual} {Interfaces}}}. \bibinfo{publisher}{ACM},
  \bibinfo{address}{Salerno Italy}, \bibinfo{pages}{1--5}.
\newblock
\showISBNx{978-1-4503-7535-1}
\urldef\tempurl%
\url{https://doi.org/10.1145/3399715.3399830}
\showDOI{\tempurl}


\bibitem[Tamminen and Poucher(2018)]%
        {tamminen_open_2018}
\bibfield{author}{\bibinfo{person}{Katherine~A. Tamminen} {and}
  \bibinfo{person}{Zoë~A. Poucher}.} \bibinfo{year}{2018}\natexlab{}.
\newblock \showarticletitle{Open science in sport and exercise psychology:
  {Review} of current approaches and considerations for qualitative inquiry}.
\newblock \bibinfo{journal}{\emph{Psychology of Sport and Exercise}}
  \bibinfo{volume}{36} (\bibinfo{year}{2018}), \bibinfo{pages}{17--28}.
\newblock
\showISSN{1878-5476}
\urldef\tempurl%
\url{https://doi.org/10.1016/j.psychsport.2017.12.010}
\showDOI{\tempurl}
\newblock
\shownote{Place: Netherlands Publisher: Elsevier Science}.


\bibitem[Tenopir et~al\mbox{.}(2019)]%
        {tenopir_seeking_2019}
\bibfield{author}{\bibinfo{person}{Carol Tenopir}, \bibinfo{person}{Lisa
  Christian}, {and} \bibinfo{person}{Jordan Kaufman}.}
  \bibinfo{year}{2019}\natexlab{}.
\newblock \showarticletitle{Seeking, {Reading}, and {Use} of {Scholarly}
  {Articles}: {An} {International} {Study} of {Perceptions} and {Behavior} of
  {Researchers}}.
\newblock \bibinfo{journal}{\emph{Publications}} \bibinfo{volume}{7},
  \bibinfo{number}{1} (\bibinfo{date}{March} \bibinfo{year}{2019}),
  \bibinfo{pages}{18}.
\newblock
\showISSN{2304-6775}
\urldef\tempurl%
\url{https://doi.org/10.3390/publications7010018}
\showDOI{\tempurl}
\newblock
\shownote{Number: 1 Publisher: Multidisciplinary Digital Publishing Institute}.


\bibitem[Wang and Zhai(2007)]%
        {wang2007learn}
\bibfield{author}{\bibinfo{person}{Xuanhui Wang} {and}
  \bibinfo{person}{ChengXiang Zhai}.} \bibinfo{year}{2007}\natexlab{}.
\newblock \showarticletitle{Learn from web search logs to organize search
  results}. In \bibinfo{booktitle}{\emph{Proceedings of the 30th annual
  international ACM SIGIR conference on Research and development in information
  retrieval}}. \bibinfo{pages}{87--94}.
\newblock


\bibitem[Wasserstein and Lazar(2016)]%
        {wasserstein2016asa}
\bibfield{author}{\bibinfo{person}{Ronald~L Wasserstein} {and}
  \bibinfo{person}{Nicole~A Lazar}.} \bibinfo{year}{2016}\natexlab{}.
\newblock \bibinfo{title}{The ASA statement on p-values: context, process, and
  purpose}.
\newblock , \bibinfo{numpages}{129--133}~pages.
\newblock


\bibitem[White(2018)]%
        {white2018opportunities}
\bibfield{author}{\bibinfo{person}{Ryen~W White}.}
  \bibinfo{year}{2018}\natexlab{}.
\newblock \showarticletitle{Opportunities and challenges in search
  interaction}.
\newblock \bibinfo{journal}{\emph{Commun. ACM}} \bibinfo{volume}{61},
  \bibinfo{number}{12} (\bibinfo{year}{2018}), \bibinfo{pages}{36--38}.
\newblock


\bibitem[White et~al\mbox{.}(2009)]%
        {white2009characterizing}
\bibfield{author}{\bibinfo{person}{Ryen~W White}, \bibinfo{person}{Susan~T
  Dumais}, {and} \bibinfo{person}{Jaime Teevan}.}
  \bibinfo{year}{2009}\natexlab{}.
\newblock \showarticletitle{Characterizing the influence of domain expertise on
  web search behavior}. In \bibinfo{booktitle}{\emph{Proceedings of the second
  ACM international conference on web search and data mining}}.
  \bibinfo{pages}{132--141}.
\newblock


\bibitem[Wu et~al\mbox{.}(2021)]%
        {wu2021predicting}
\bibfield{author}{\bibinfo{person}{Jian Wu}, \bibinfo{person}{Rajal Nivargi},
  \bibinfo{person}{Sree Sai~Teja Lanka}, \bibinfo{person}{Arjun~Manoj Menon},
  \bibinfo{person}{Sai~Ajay Modukuri}, \bibinfo{person}{Nishanth Nakshatri},
  \bibinfo{person}{Xin Wei}, \bibinfo{person}{Zhuoer Wang},
  \bibinfo{person}{James Caverlee}, \bibinfo{person}{Sarah~M Rajtmajer},
  {et~al\mbox{.}}} \bibinfo{year}{2021}\natexlab{}.
\newblock \showarticletitle{Predicting the Reproducibility of Social and
  Behavioral Science Papers Using Supervised Learning Models}.
\newblock \bibinfo{journal}{\emph{arXiv preprint arXiv:2104.04580}}
  (\bibinfo{year}{2021}).
\newblock


\bibitem[Yan and Ding(2010)]%
        {yan_weighted_2010}
\bibfield{author}{\bibinfo{person}{Erjia Yan} {and} \bibinfo{person}{Ying
  Ding}.} \bibinfo{year}{2010}\natexlab{}.
\newblock \showarticletitle{Weighted citation: {An} indicator of an article's
  prestige}.
\newblock \bibinfo{journal}{\emph{Journal of the American Society for
  Information Science and Technology}} \bibinfo{volume}{61},
  \bibinfo{number}{8} (\bibinfo{year}{2010}), \bibinfo{pages}{1635--1643}.
\newblock
\newblock
\shownote{ISBN: 1532-2882 Publisher: Wiley Online Library}.


\bibitem[Yang et~al\mbox{.}(2020)]%
        {yang2020estimating}
\bibfield{author}{\bibinfo{person}{Yang Yang}, \bibinfo{person}{Wu Youyou},
  {and} \bibinfo{person}{Brian Uzzi}.} \bibinfo{year}{2020}\natexlab{}.
\newblock \showarticletitle{Estimating the deep replicability of scientific
  findings using human and artificial intelligence}.
\newblock \bibinfo{journal}{\emph{Proceedings of the National Academy of
  Sciences}} \bibinfo{volume}{117}, \bibinfo{number}{20}
  (\bibinfo{year}{2020}), \bibinfo{pages}{10762--10768}.
\newblock


\bibitem[Zhu et~al\mbox{.}(2021)]%
        {zhu_recommending_2021}
\bibfield{author}{\bibinfo{person}{Yifan Zhu}, \bibinfo{person}{Qika Lin},
  \bibinfo{person}{Hao Lu}, \bibinfo{person}{Kaize Shi}, \bibinfo{person}{Ping
  Qiu}, {and} \bibinfo{person}{Zhendong Niu}.} \bibinfo{year}{2021}\natexlab{}.
\newblock \showarticletitle{Recommending scientific paper via heterogeneous
  knowledge embedding based attentive recurrent neural networks}.
\newblock \bibinfo{journal}{\emph{Knowledge-Based Systems}}
  \bibinfo{volume}{215} (\bibinfo{date}{March} \bibinfo{year}{2021}),
  \bibinfo{pages}{106744}.
\newblock
\showISSN{0950-7051}
\urldef\tempurl%
\url{https://doi.org/10.1016/j.knosys.2021.106744}
\showDOI{\tempurl}


\end{thebibliography}

\end{document}